\documentclass[journal,twoside,web]{ieeecolor}
\usepackage{tmi}
\usepackage{cite}
\usepackage{amsmath,amssymb,amsfonts}
\usepackage{algorithmic}
\usepackage{graphicx}
\usepackage{textcomp}
\usepackage{multirow}
\usepackage{url}
\usepackage[running]{lineno}
\usepackage{comment}
\usepackage{xcolor}

\makeatletter
\renewcommand{\maketag@@@}[1]{\hbox{\m@th\normalsize\normalfont#1}}%
\makeatother

\newtheorem{theorem}{Theorem}
\def\BibTeX{{\rm B\kern-.05em{\sc i\kern-.025em b}\kern-.08em
    T\kern-.1667em\lower.7ex\hbox{E}\kern-.125emX}}
\def \gaor {\textcolor{black}} 
\def \gaog {\textcolor{black}} 
\def \gaom {\textcolor{black}} 
\def \gaob {\textcolor{black}} 
\def \gaoc {\textcolor{black}}

\begin{document}
\title{Hierarchical Perception Adversarial Learning Framework for Compressed Sensing MRI}
\author{Zhifan Gao, \IEEEmembership{Member, IEEE}, Yifeng Guo, Jiajing Zhang, Tieyong Zeng, Guang Yang, \IEEEmembership{Senior Member, IEEE}
\thanks{This work was supported by National Key R\&D Program of China (2022YFE0209800), National Natural Science Foundation of China (U1908211, 62101606), Shenzhen Science and Technology Program (Grant No. GXWD20201231165807008, 20200825113400001), the ERC IMI (101005122), the H2020 (952172), the MRC (MC/PC/21013), the Royal Society (IEC/NSFC/211235), the NVIDIA Academic Hardware Grant Program, NIHR Imperial Biomedical Research Centre (RDA01), Imperial–Nanyang Technological University Collaboration Fund, UKRI MRC with MSIT and NRF Fund, and the UKRI Future Leaders Fellowship (MR/V023799/1), Grant NSFC/RGC N\_CUHK 415/19, Grant ITF MHP/038/20, Grant CRF 8730063, Grant RGC 14300219, 14302920, 14301121.
(Corresponding author: Guang Yang.)}
\thanks{Z. Gao, Y. Guo and J. Zhang are with School of Biomedical Engineering, Sun Yat-sen University, Shenzhen, China  (e-mail: gaozhifan@mail.sysu.edu.cn; guoyf25@mail2.sysu.edu.cn; zhangjj83@mail2.sysu.edu.cn).}
\thanks{T. Zeng is with the Department of Mathematics, The Chinese University of
Hong Kong, Shatin, Hong Kong (e-mail: zeng@math.cuhk.edu.hk). }
\thanks{G. Yang is with Cardiovascular Research Centre, Royal Brompton Hospital,
UK and also with National Heart \& Lung Institute, Imperial College London, London, UK (e-mail: g.yang@imperial.ac.uk). }
}

\maketitle

\begin{abstract}

The long acquisition time has limited the accessibility of magnetic resonance imaging (MRI) because it leads to patient discomfort and motion \gaob{artifacts}. Although several MRI techniques have been proposed to reduce the acquisition time, compressed sensing in magnetic resonance imaging (CS-MRI) enables fast acquisition without compromising SNR and resolution. However, \gaob{existing} CS-MRI methods suffer from the challenge of aliasing \gaob{artifacts}. This challenge results in the noise-like \gaob{textures} and \gaob{missing the fine details}, \gaob{thus leading} to \gaob{unsatisfactory} reconstruction performance. To tackle this challenge, we propose \gaob{a} hierarchical perception adversarial learning framework (HP-ALF). HP-ALF can perceive the image information in the hierarchical mechanism: image-level perception and patch-level perception. The former can reduce the visual perception difference in the entire image, and thus achieve \gaob{aliasing artifact removal}. The latter can reduce this difference in the regions of the image, and thus recover \gaob{fine details}. Specifically, HP-ALF achieves the hierarchical mechanism by \gaob{utilizing multilevel} perspective discrimination. This discrimination can provide the information from two perspectives (overall and regional) for adversarial learning. It also \gaob{utilizes} a global and local coherent discriminator to provide structure information to the generator during training. In addition, HP-ALF contains a context-aware learning block to effectively exploit the slice information between individual images for better reconstruction performance. The experiments validated on three datasets demonstrate the effectiveness of HP-ALF and its superiority to the comparative methods.
\end{abstract}


\section{Introduction}
\label{sec:introduction}

\IEEEPARstart{T}{he} long acquisition time \gaoc{in} magnetic resonance imaging (MRI) \gaob{limits} the accessibility of this modality \cite{jaspan2015compressed}. It can lead to \gaob{patient discomfort and motion artifacts} \cite{mann2016accelerating,zhou2018simple}. Several \gaoc{existing} MRI techniques have been proposed to reduce the acquisition time, such as parallel imaging (PI), simultaneous multislice (SMS) and compressed sensing in magnetic resonance imaging (CS-MRI). \gaoc{CS-MRI enables \gaob{a} significant reduction \gaob{in} the MRI acquisition time} without compromising SNR and resolution \gaoc{with respect to} other MRI techniques \cite{lustig2008compressed}. \gaoc{It can avoid the deterioration of image quality because it performs \gaob{nonlinear optimization} on highly-undersampled raw data}. \gaoc{Thus, CS-MRI allows} clinicians to complete \gaob{multidimensional} scans in \gaoc{a} clinically feasible scan time. For example, CS-MRI can reduce the scan time by 29.3\% in a daily clinical routine study for the brain \cite{monch2020magnetic} and reduce the time for volumetric cardiac-resolved flow imaging \gaob{sequences} (4D flow) from an hour to $5–10$ \gaoc{minutes} \cite{cheng2016comprehensive}.

\begin{figure}
\centering
	\includegraphics[width=0.49\textwidth]{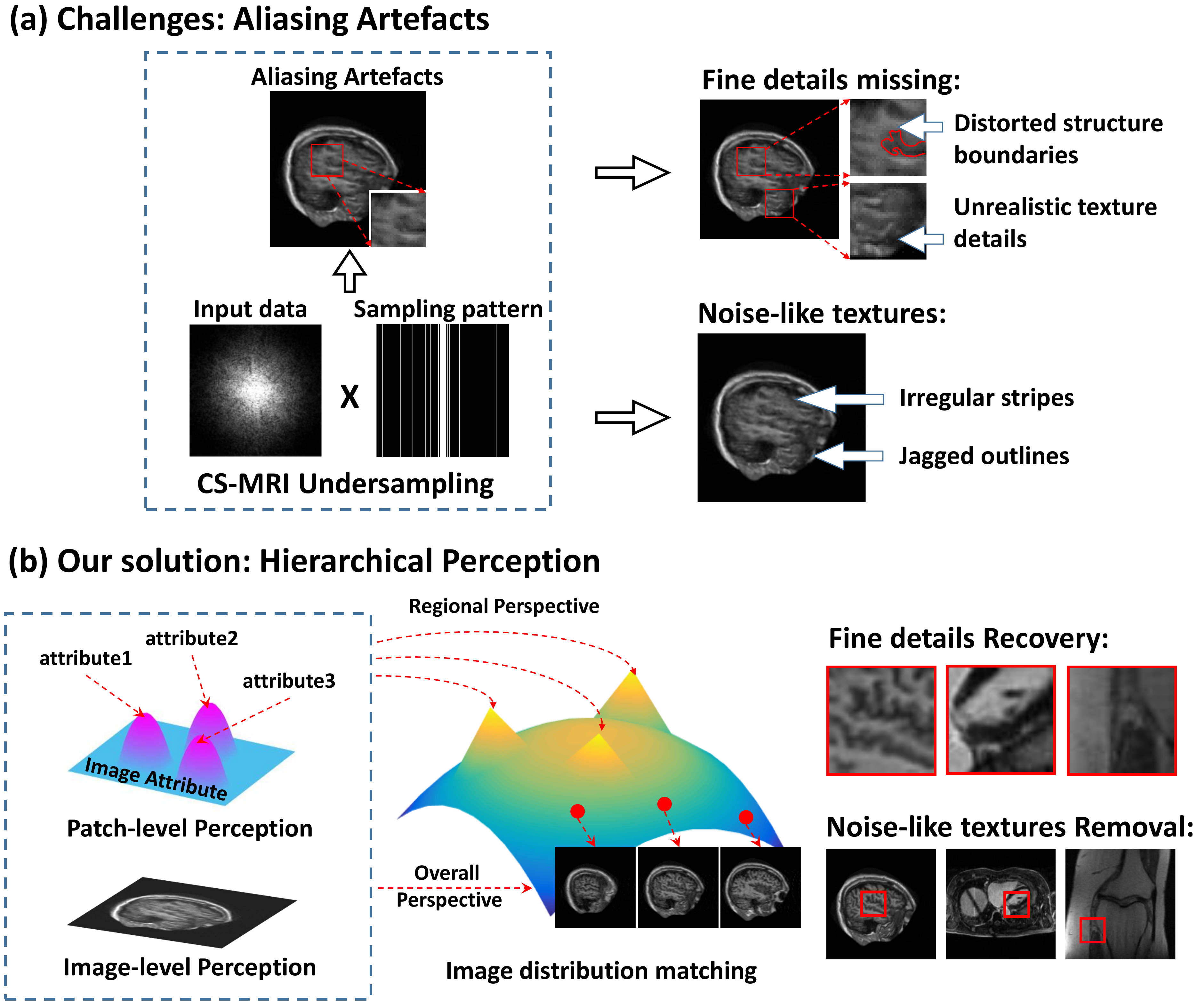}
\caption{The contribution of our hierarchical perception adversarial learning framework (HP-ALF). HP-ALF \gaob{utilizes} hierarchical perception to address the challenge of aliasing \gaob{artifacts}. This challenge is derived from the \gaob{high undersampling in} CS-MRI, which applies \gaob{the} sampling patterns to the input data. It results \gaob{in noise-like texture and missing fine details} in the image domain. Hierarchical perception includes image-level perception and patch-level perception. They can reduce the visual perception difference from overall perspective and regional perspective, \gaob{and thus remove noise-like texture and restore fine details}.}
\label{Figure1}
\end{figure}

However, CS-MRI still \gaoc{has} unsatisfactory reconstruction performance because of aliasing \gaob{artifacts} \cite{donoho2006compressed}. This challenge leads to \gaob{difficulty in applying} CS-MRI in the examination of some clinical indications, such as epilepsy and pediatric anesthesia \cite{delattre2020compressed,robertson2018imaging}. The aliasing \gaob{artifacts} are derived from the \gaob{high undersampling} \gaoc{in} CS-MRI \cite{lustig2008compressed}. They cause \gaob{noise-like textures} and the \gaob{missing fine details} to corrupt the \gaoc{reconstructed image} \cite{lustig2007sparse}, \gaoc{\gaob{as} shown in Figure \ref{Figure1}(a)}. First, \gaob{noise}-like texture refers to \gaob{an} irregular pattern that blurs the image globally \cite{lustig2008compressed}. It obscures and weakens the \gaoc{appearance of the structure and the edge}, and thus interferes with the extraction of \gaob{this} feature information. \gaoc{This} interference leads to \gaob{distortion} of the reconstructed image. Second, the fine \gaoc{detail} missing refers to the blurry textures in the different parts of the image \cite{yang2017dagan,mardani2018deep}. It conceals and weakens the structure \gaoc{boundary and the} texture details \gaoc{and results} in \gaoc{the} unrealistic appearance of \gaob{small} structures. The unrealistic image appearance reduces the perceptual quality of the reconstructed image.

Existing CS-MRI methods \gaob{have difficulty addressing} the challenge of aliasing \gaob{artifacts} \cite{chen2022ai,block2007undersampled,yang2017dagan}. Although they perform well under the interference of noise-like texture \cite{jung2009k,feng2021task}, \gaob{aliasing} artefacts still \gaob{come} from the \gaob{missing fine details} \cite{mardani2018deep,shaul2020subsampled}. This is because these methods perceive the visual difference between the reconstructed image and original image in the overall perspective (for removing noise-like texture), rather than \gaoc{in} the regional perspective (for restoring the fine details).
First, \gaob{conventional} CS-MRI methods \gaoc{tend to} focus on \gaob{low-frequency} image information owing to \gaob{hand-crafted} feature extraction rather than \gaob{high}-frequency information \gaoc{corresponding to} \gaob{fine} details \cite{block2007undersampled,jung2009k}.
This corrupts the reconstruction of the fine details.
Second, the existing deep-learning-based CS-MRI methods also face this difficulty, although they enable \gaob{both} high- and low-frequency information extraction \cite{qin2018convolutional,feng2021task,yanga}. \gaoc{They usually rely} on the computation of the pixel-wise distance in \gaob{the} spatial or frequency domain, \gaoc{and thus have to} smooth the fine details in the reconstructed image \cite{blau2018perception,yang2017dagan,mardani2018deep}. This may lead to \gaob{difficulty in reducing} the visual perception difference between \gaob{fine} details.
Third, \gaob{adversarial} learning methods can reconstruct the image with \gaoc{fine details} by bringing in the distribution distance \cite{goodfellow2014generative,yang2017dagan,mardani2018deep,quan2018compressed, Xiangli*2020Real}. However, these methods recover \gaob{fine} details from the overall perspective. This is because the information of these details \gaob{is} related to the \gaob{attributes} in different image regions. These attributes represent the parts of the image, \gaoc{including} \gaoc{the} different levels of aliasing \gaob{artifacts}. The level of aliasing \gaob{artifacts} changes in different image regions \gaoc{and influences the attributes in} the reconstruction of \gaob{fine} details.  Therefore, the visual perception difference needs to be measured not only from the overall perspective but also from the regional perspective.

In this paper, we propose \gaoc{the} hierarchical perception adversarial learning framework (HP-ALF) to tackle the challenge of aliasing \gaob{artifacts}, \gaob{as} shown \gaoc{in} Figure \ref{Figure1}(b). \gaoc{It builds the perception of image information by the hierarchical mechanism}: image-level and patch-level.
First, \gaob{image}-level perception refers to the extraction of the image information from the overall perspective. It \gaoc{perceives and calculates} the global aliasing \gaob{artifacts}, and thus \gaob{reduces} the visual perception difference in the entire image.
Second, \gaob{patch}-level perception refers to the extraction of \gaob{detailed} information from the regional perspective. It perceives and calculates the local aliasing \gaob{artifacts} and thus reduces the visual perception difference in the different regions of the image.
Thus, HP-ALF not only builds the image-level perception to reduce the visual perception difference from the overall perspective, but also \gaoc{enables} the patch-level perception from the regional perspective. It can remove noise-like \gaob{textures} and restore the fine details simultaneously, \gaob{thus reducing} image distortion and \gaob{improving} perceptual quality.
Specifically, HP-ALF is implemented by the \gaob{multilevel} perspective discrimination. It matches the distributions of the reconstructed image and the original image by comparing their quality \gaoc{difference} from both the overall and regional perspectives. Therefore, it facilitates \gaoc{the reconstruction of \gaob{high}-quality images}.

Our contributions can be \gaob{summarized} as follows:

1. We develop a CS-MRI framework to reconstruct \gaoc{\gaob{high}-quality MRI images}. It enables the hierarchical perception \gaoc{by} \gaob{multilevel} perspective discrimination to reduce the visual perception difference from the overall perspective and \gaoc{the} regional \gaob{perspectives}. This \gaoc{framework} can achieve both the noise-like texture removal and \gaob{fine} \gaoc{detail} restoration to address the aliasing \gaob{artifacts} owing to the \gaob{high} undersampling.

2. We design a \gaob{minimization} problem for CS-MRI to evaluate the perceptual quality from the \gaob{overall and regional perspectives}. To solve this problem, we extract the global and local structure information as well as the slice information in the image sequence. First, we \gaoc{build} a global and local coherent discriminator to provide \gaoc{the} detailed per-pixel decision to the generator while maintaining the global coherence of \gaoc{the} reconstructed images. Then, the context-aware learning block in the generator exploits the slice information from the \gaoc{MRI sequence}.

3. We validate \gaoc{our framework} on three datasets for different anatomical structures (brain, heart, knee). The experimental results demonstrate the effectiveness of \gaoc{our framework}, as well as its superiority to \gaob{comparative} CS-MRI methods.

This work advances our preliminary work in MICCAI 2020 \cite{guo2020deep}. First, it extends the \gaob{minimization} problem that is applied to the noise-like texture removal and \gaob{restores} the fine details simultaneously. Second, it designs a novel objective function that transforms the single-level perspective discrimination into the \gaob{multilevel} perspective discrimination for perceptual quality improvement. Third, it extends the discriminator to a \gaob{U-net-based} architecture that can reconstruct globally and locally coherent images for fine \gaoc{detail} preservation. \gaoc{Finally}, the experiments are extended to three datasets imaging different organs and three more validations (\gaob{evaluating} the effectiveness of the proposed objective function, the \gaob{U-net-based} discriminator, and the context-aware learning block).

\section{Related Works}
The current CS-MRI methods can be broadly classified into conventional CS-MRI methods, traditional deep \gaob{learning-based} CS-MRI methods, and adversarial learning based CS-MRI methods. However, these methods ignore the restoration of \gaob{fine} details.

First, \gaob{conventional} CS-MRI methods \gaoc{have the challenge} \gaob{of extracting} high-frequency information \gaoc{for feature representation of \gaob{fine} details}.
These methods include sparsity-based and dictionary learning-based CS-MRI. First, \gaoc{sparsity-based} CS-MRI reconstruction methods have been developed to leverage the sparsity of \gaob{signals} by using predefined and fixed sparse transformations \cite{lustig2008compressed,zengb} and \gaob{exploiting spatiotemporal} correlations \cite{jung2009k}. These methods usually rely on the experience-based determination of what kind of low-frequency image information is helpful. Second, compared to \gaob{sparsity}-based methods, the dictionary learning (DL) can generate data-specific dictionaries and improve image quality \cite{aharon2006k,zenga}. However, the dictionary learning method \gaoc{has difficulty reconstructing} high-frequency features because these dictionaries are still designed based on \gaob{low}-frequency image features.

\gaob{In addition}, existing CS-MRI methods \gaob{still difficulty reconstructing} realistic fine details, although deep learning \gaoc{shows} high potential in \gaoc{many medical image applications} \cite{gaoa,Forea,guoa,Foreb,gaob}. Compared with \gaob{conventional} CS-MRI methods, \gaob{existing} \gaob{deep learning-based} methods \gaoc{can extract the high- and low-frequency features}.
These methods can be divided into two classes \cite{chen2022ai}.
First, \gaob{end-to-end optimization} models the inverse acquisition to achieve the fast MRI reconstruction. For example, Feng et al. \cite{feng2021task} introduced an end-to-end task transformer network (T2Net) \gaob{utilizing} an $\ell_{1}$ loss function. Second, \gaob{unrolled optimization} incorporates \gaob{prior} domain knowledge about the expected \gaob{properties} of MR images. For example, Qin et al. \cite{qin2018convolutional} proposed a convolutional recurrent neural network (CRNN) method based on \gaob{unrolled optimization} with the pixel-wise distance in the spatial domain. Guo et al. \cite{guo2021over} proposed an unrolled model based on the novel convolutional recurrent neural network (OUCR) with $\ell_{1}$ loss. Hu et al. \cite{hu2021self} proposed a self-supervised unrolled model (SSL-MRI) based on the parallel network training framework with a pixel-wise loss.
However, these methods fail to reconstruct the realistic fine details. This is because \gaob{it is} difficult to improve the perceptual quality of \gaob{fine} details by reducing the visual perception difference. The visual perception difference \gaoc{is associated with} the distance between the reconstructed image distribution and the original image distribution \cite{blau2018perception}. \gaoc{These} deep-learning-based methods \gaob{have difficulty calculating} this distribution distance \cite{blau2018perception}.

\gaob{Furthermore}, \gaoc{adversarial learning} can introduce the distribution distance to further reduce the visual perception difference \cite{goodfellow2014generative,yang2017dagan,mardani2018deep,quan2018compressed,wua}. However, it is still unsatisfactory \gaob{for recovering} fine details. \gaob{Adversarial} learning methods \gaoc{enable the computation} of the distribution distance \cite{blau2018perception}. \gaoc{Thus}, these methods can capture and reduce visual perception \gaob{differences \gaoc{to improve}} the perceptual quality of the reconstructed image. \gaoc{The existing methods} can be divided into two classes. First, the loss-variant model \gaob{utilizes} auxiliary penalties to improve the perceptual quality of the reconstructed image.
For instance, Yang et al. \cite{yang2017dagan} proposed a combination of pixel-wise, perceptual, and GAN losses to achieve fast CS-MRI \gaob{utilizing} the conditional \gaob{generative adversarial network}-based model (DAGAN). Mardani et al. \cite{mardani2018deep} proposed a mixture of pixel-wise and least-squares GAN (GANCS) losses in which the least-squares GAN learns the texture details and the pixel-wise loss suppresses high-frequency noise.
Second, the architecture-variant model \gaoc{relies on} the MRI data \gaoc{characteristics} to improve the reconstruction performance. For example, Shaul et al. \cite{shaul2020subsampled} proposed a two-stage GAN framework (Sub-GAN) including a cascade of a \textit{k}-space and an image-space \gaob{U-Net} with a mixture loss. Korkmaz et al. \cite{korkmaz2022unsupervised} proposed a novel unsupervised MRI reconstruction based on an unconditional deep adversarial network (SLATER) \gaob{utilizing} a mixture loss. Wei et al. \cite{wei2022undersampled} introduced a two-stage generative adversarial network utilising cross-domain learning with $\ell_{1}$ and $\ell_{2}$ pixel-wise loss.
However, these methods \gaob{have difficulty improving} the perceptual quality of fine details. This is because they reduce the visual perception difference only from the overall perspective and thus \gaob{lead} to the unrealistic reconstruction of the fine details.

\begin{figure*}
\centering
	\includegraphics[width=0.93\textwidth]{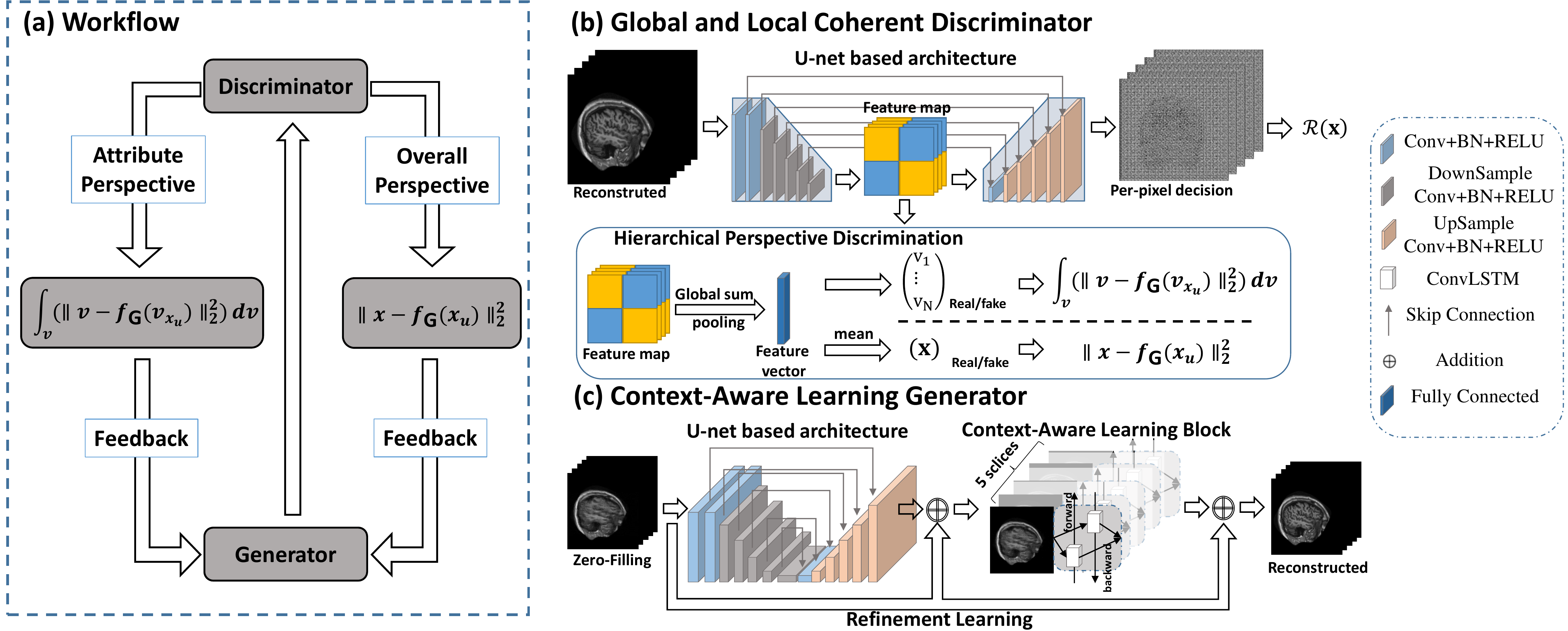}
\caption{Overview of our HP-ALF. (a) HP-ALF \gaob{utilizes multilevel} perspective discrimination to achieve the hierarchical perception by providing information from the overall perspective and \gaob{the} regional perspective. (b) Our global and local coherent discriminator \gaob{utilizes a U-net-based} architecture to provide fine details information during training. It is achieved by using the decoder of U-net to provide a detailed per-pixel decision to the generator. (c) Context-aware learning generator includes a \gaob{U-net-based} architecture and a context-aware learning block. U-net can \gaob{utilize} 2D spatial information while the context-aware learning block can exploit 3D spatial feature from the sequential MRI data.}
\label{Figure_model}
\end{figure*}

\section{Method}
\subsection{Problem Statement}
Let $\mathbf{x} \in \mathbb{C}^{N}$ \gaoc{be} the slice of 2D images to be reconstructed, where each slice consists of $\sqrt{N} \times \sqrt{N}$ pixels for one image. The problem is to reconstruct $\mathbf{x}$ from $\mathbf{y} \in \mathbb{C}^{M}$ ($M \ll N$), undersampled measurements in \textit{k}-space, such that $\mathbf{y}=\mathbf{F}_{u} \mathbf{x}+\epsilon$. $\mathbf{F}_{u}$ is the undersampling Fourier encoding operator and $\epsilon$ is complex Gaussian noise \cite{donoho2006compressed}. However, such measurements are underdetermined even in the absence of noise because of the violation of the Nyquist-Shannon sampling theorem \cite{lustig2008compressed}. Therefore, the corresponding linear inversion for CS-MRI $\mathbf{x}_{u}=\mathbf{F}_{u}^{H}\mathbf{y}$ is usually ill-posed, \gaog{where $H$ denotes the Hermitian transpose operation.} $\mathbf{x}_{u}$ suffers from the challenge of aliasing \gaob{artifacts}. \gaoc{The} \gaob{artifacts} result in \gaob{noise-like textures} and \gaob{missing fine details} in the image domain. The existing CS-MRI methods address the challenge by formulating a \gaob{minimization} problem \cite{lustig2008compressed,yang2017dagan}:

\begin{equation}
\min _{\mathbf{x}} \lambda\left\|\mathbf{y}-\mathbf{F}_{u} \mathbf{x}\right\|_{2}^{2}+\mathcal{R}(\mathbf{x}),
\label{ConventionCS}
\end{equation}
\gaoc{where} $\left\|\mathbf{y}-\mathbf{F}_{u}\mathbf{x}\right\|_{2}^{2}$ is the data fidelity term \cite{donoho2006compressed} and $\mathcal{R}(\mathbf{x})$ is the \gaob{regularization} term. $\lambda$ is the \gaob{regularization} parameter. However, these methods \gaoc{have difficulty removing noise-like textures and restoring fine details} simultaneously. This is because Equation (\ref{ConventionCS}) tends to \gaob{realize} noise-like texture removal but \gaob{neglects fine detail} recovery. The detailed image information is related to the \gaob{attributes} in different image regions. These methods can only perceive the image from the overall perspective. Therefore, CS-MRI reconstruction requires both noise-like texture removal and fine detail recovery from different perspectives. HP-ALF \gaoc{builds} an additional \gaob{regularization} term (i.e., \gaob{a} hierarchical perception term) in Equation (\ref{ConventionCS}). This term can reduce the visual perception difference from the overall perspective and the regional perspective. The CS-MRI reconstruction problem can be reformulated as a different \gaob{minimization} problem:

\begin{equation}
\begin{gathered}
\min _{\mathbf{x},\mathbf{v}} \int_{\mathbf{v}}\left(\left\|\mathbf{v}_{\mathbf{x}}-f_{\text {G}}\left(\mathbf{v}_{\mathbf{x}_{u}}\right)\right\|_{2}^{2}\right) d \mathbf{v}+\|\mathbf{x}-f_{\text {G}}\left(\mathbf{x}_{u}\right)\|_{2}^{2} \\
+\lambda\left\|\mathbf{y}-\mathbf{F}_{u} \mathbf{x}\right\|_{2}^{2}+\mathcal{R}(\mathbf{x}), \\
\gaor{
\mathcal{R}(\mathbf{x})=\sum_{i,j}\left\|\left[f_\text{D}\left(\mathbf{x}\right)\right]_{i,j}-f_\text{D}\left(f_\text{G}\left(\mathbf{x}_{u}\right)\right)]_{i,j}\right\|
}
\end{gathered}
\label{SolutionCS}
\end{equation}
\gaoc{where} $f_{\text {G}}: \mathbb{C}^{N} \mapsto \mathbb{C}^{N}$ is the model that reconstructs images from $\mathbf{x}_{u}$ to address the challenge of aliasing \gaob{artifacts}. $f_{\text {D}}$ represents the model that outputs local (per-pixel) changes between the reconstructed image and the original image. $\mathbf{x}_{u}$ \gaob{represents} the $n$ slices \gaob{of} sequential MRI images. Further explanations of the variable $\mathbf{v}$, hierarchical perception term and \gaob{regularization} term are as follows.

\textbf{The variable $\mathbf{v}$.} The variable $\mathbf{v}$ is the mathematical symbol of the attribute. The attribute represents the parts of the image with different levels of aliasing \gaob{artifacts}. \gaoc{Thus, $\mathbf{v}_{\mathbf{x}}$ \gaob{represents} the parts of the reconstructed image $\mathbf{x}$ with different levels of aliasing \gaob{artifacts}. $\mathbf{v}_{\mathbf{x}_{u}}$ \gaob{represents} those of the zero-filled reconstruction $\mathbf{x}_{u}$.} \gaoc{Then, the term $\int_{\mathbf{v}}\left(\left\|\mathbf{v}_{\mathbf{x}}-f_{\text {G}}\left(\mathbf{v}_{\mathbf{x}_{u}}\right)\right\|_{2}^{2}\right) d \mathbf{v}$ in Equation (\ref{SolutionCS}) \gaob{aims} to} compute the difference \gaob{in} the attributes between the reconstructed image and the original image. \gaob{This} facilitates Equation (\ref{SolutionCS}) \gaob{in focusing} on the removal of local aliasing \gaob{artifacts} in different regions. Specifically, Equation (\ref{SolutionCS}) \gaob{utilizes} the $\ell_{2}$ distance to calculate the difference of the variable $\mathbf{v}$. Then, it sums up all the difference results to obtain the integral result of $\int_{\mathbf{v}}\left(\left\|\mathbf{v}_{\mathbf{x}}-f_{\text {G}}\left(\mathbf{v}_{\mathbf{x}_{u}}\right)\right\|_{2}^{2}\right) d \mathbf{v}$.

\textbf{Hierarchical perception term.}
This term is the additional \gaob{regularization} term in Equation (\ref{ConventionCS}), including two parts. First, $\int_{\mathbf{v}}\left(\left\|\mathbf{v}_{\mathbf{x}}-f_{\text {G}}\left(\mathbf{v}_{\mathbf{x}_{u}}\right)\right\|_{2}^{2}\right) d \mathbf{v}$ represents the difference \gaob{in} the local aliasing \gaob{artifact} between the reconstructed image and the original image. Therefore, it can calculate and reduce the visual perception difference in the different regions of the image and thus recover fine details. Second, $\|\mathbf{x}-f_{\text {G}}\left(\mathbf{x}_{u}\right)\|_{2}^{2}$ represents the difference \gaob{in} the global aliasing \gaob{artifact} between the reconstructed image and the original image. Therefore, it can calculate and reduce the visual perception difference in the entire image, and thus achieve the aliasing \gaob{artifact} removal.

\textbf{Regularization term.} The term $\mathcal{R}(\mathbf{x})$ in Equation (\ref{SolutionCS}) represents the local difference of the images at pixel $(i,j)$ calculated by the model $f_\text{D}$. It provides the detailed per-pixel decision during the \gaob{optimization} process while maintaining the global coherence of \gaoc{the} reconstructed images.

Our HP-ALF achieves Equation (\ref{SolutionCS}) \gaob{by image}-level perception and the patch-level perception. \gaob{Image}-level perception and \gaob{patch}-level perception \gaob{are represented by} the first term and the second term in Equation (\ref{SolutionCS}), respectively. They can reduce the visual perception difference from the overall perspective and the regional perspective. Specifically, HP-ALF \gaoc{constructs} the encoder of the global and local coherent discriminator to achieve \gaob{image}-level perception and \gaob{patch}-level perception. The decoder of this discriminator can preserve \gaoc{the fine} details. Then, a context-aware learning block in the generator exploits the slice information. In addition, the loss function in HP-ALF can be \gaob{optimized} for Equation (\ref{SolutionCS}). Figure \ref{Figure_model} shows the details of the HP-ALF.

\subsection{Multilevel Perspective Discrimination}\label{sec:method RealnessGAN}
\gaog{We \gaob{propose multilevel} perspective discrimination in HP-ALF to achieve the patch-level perception and image-level perception in Equation (\ref{SolutionCS}), as shown in Figure \ref{Figure_model}(a).}
It is \gaor{built by the objective function of the} generative adversarial network (GAN) including \gaob{image}-level perspective discrimination and \gaob{patch}-level discrimination.
First, the patch-level perspective discrimination refers \gaob{to} $\int_{\mathbf{v}}\left(\left\|\mathbf{v}_{\mathbf{x}}-f_{\text {G}}\left(\mathbf{v}_{\mathbf{x}_{u}}\right)\right\|_{2}^{2}\right) d \mathbf{v}$ in Equation (\ref{SolutionCS}). It measures the difference \gaob{in} $\mathbf{v}$ between the reconstructed image and the original image. Therefore, it can provide information from the regional perspective for distribution matching in adversarial learning.
Second, \gaob{image}-level perspective discrimination refers \gaob{to} $\|\mathbf{x}-f_{\text {G}}\left(\mathbf{x}_{u}\right)\|_{2}^{2}$ in Equation (\ref{SolutionCS}). It measures the difference between the reconstructed image and the original image. Therefore, it can provide information from the overall perspective for distribution matching in adversarial learning.
The encoder of the discriminator in HP-ALF is constructed based on $\|\mathbf{x}-f_{\text {G}}\left(\mathbf{x}_{u}\right)\|_{2}^{2}$ and $\int_{\mathbf{v}}\left(\left\|\mathbf{v}_{\mathbf{x}}-f_{\text {G}}\left(\mathbf{v}_{\mathbf{x}_{u}}\right)\right\|_{2}^{2}\right) d \mathbf{v}$. The \gaob{optimization} process of the encoder is presented as a two-player min-max value function. \gaoc{This value function includes the} discriminator loss and the generator loss for the model training.

\begin{figure}[t!]
\centering
	\includegraphics[width=0.49\textwidth]{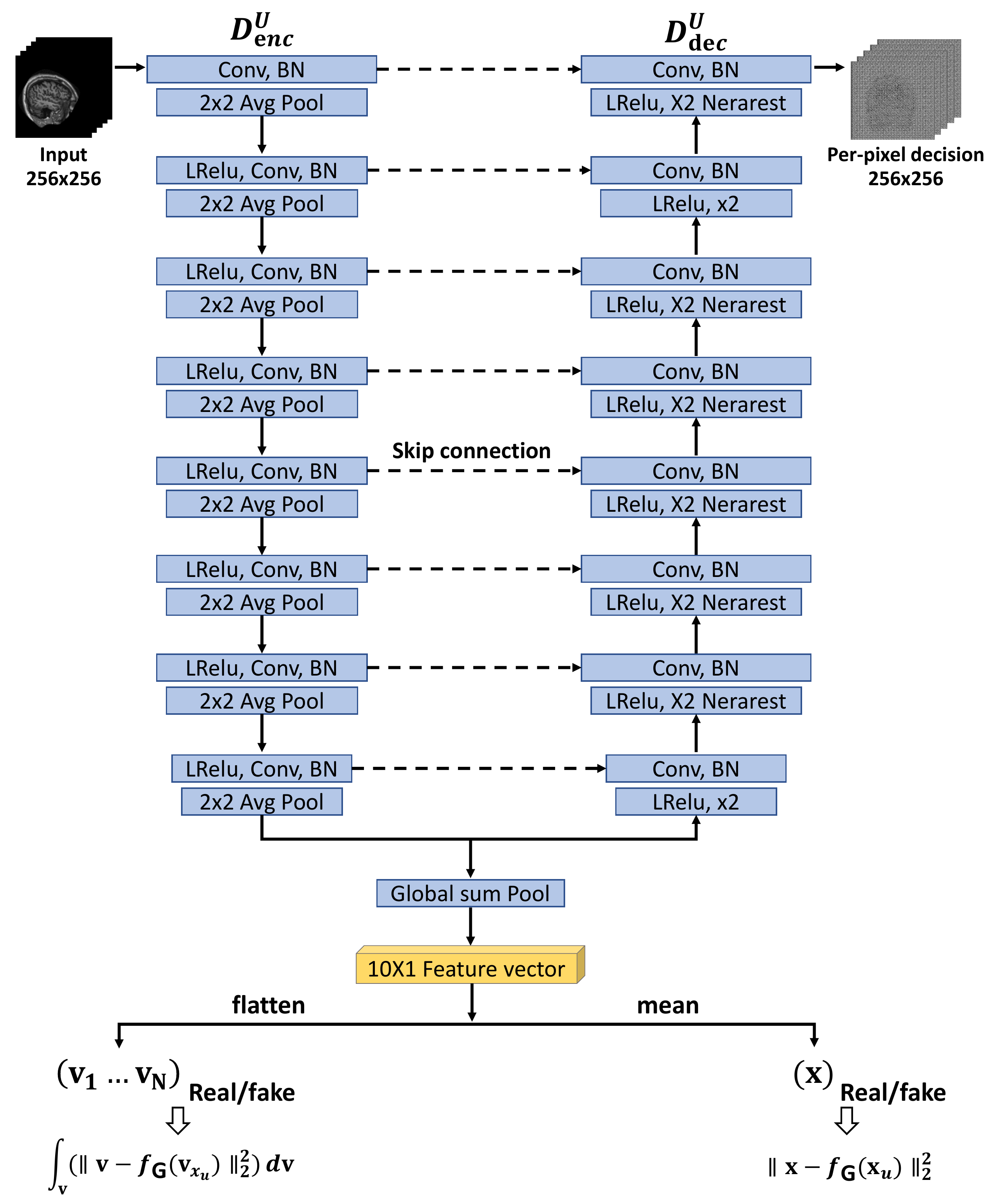}
\caption{\gaob{The network details of the global and local coherent discriminator. The encoder in our discriminator represents the features and downsamples it in each level. The decoder in our discriminator represents the features and then upsamples it in each level. This discriminator also performs the multi-level discrimination for the reconstructed images.}}
\label{Figure25}
\end{figure}

\gaom{
Specifically, the output of the encoder in the discriminator is applied as the input of this discrimination. The output of the encoder is divided into two parts. The first part is a scalar as that in the traditional adversarial learning method \cite{goodfellow2014generative}. It is the mean value of the feature map obtained from the last layer \gaob{of} the encoder. This scalar represents the difference between the entire reconstructed image and the entire original image. Therefore, it can be used to achieve \gaob{image}-level perspective discrimination.
The second part is the feature map obtained from the last layer in the encoder. It can be flattened as a discrete distribution $p_{\text{perspective}}$.}
This distribution is constructed based on the variable $\mathbf{v}$. Each element in this distribution corresponds to an image attribute. Therefore, this distribution can be \gaoc{applied for} patch-level perspective discrimination. For the input sample $\boldsymbol{x}$, the corresponding discriminator output can be represented as $D(\boldsymbol{x})=\left\{p_{\text {perspective }}(\boldsymbol{x}, \mathbf{v}) ; \mathbf{v} \in \Omega\right\}$, where $\Omega$ is the set of outcomes of $p_{\text {perspective }}$. Each outcome $\mathbf{v}$ corresponds to the attribute in different regions. \gaob{To} measure the distance between distributions, HP-ALF \gaoc{constructs} two normal distributions with a positive skew and a negative skew as $\mathcal{R}_{1}$ (real) and $\mathcal{R}_{0}$ (fake), \gaob{respectively, similar to} the virtual ground-truth scalars 0 and 1 in \gaoc{the} standard GAN. \gaob{$\mathcal{R}_{1}$} and $\mathcal{R}_{0}$ are also defined on $\Omega$. Accordingly, the \gaob{JS divergence} in the standard GAN is replaced with the Kullback-Leibler (KL) divergence. The min-max game between $G$ and $D$ thus becomes:
\begin{equation}\label{RealnessGAN}
\begin{aligned}
&\max _{G} \min _{D} V(G, D)= \\
&\mathbb{E}_{\boldsymbol{x} \sim p_{\text {data }}}\left[\mathcal{D}_{\mathrm{KL}}\left(\mathcal{R}_{1}(\mathbf{v}) \| D(\boldsymbol{x})\right)+\log (D(\boldsymbol{x}))\right] \\
&+\mathbb{E}_{\boldsymbol{x} \sim p_{g}}\left[\mathcal{D}_{\mathrm{KL}}\left(\mathcal{R}_{0}(\mathbf{v}) \| D(\boldsymbol{x})\right)+\log (1-D(\boldsymbol{x}))\right].
\end{aligned}
\end{equation}
\gaoc{where} $D$ in Equation (\ref{RealnessGAN}) corresponds to the encoder of the discriminator. It aims to maximize the value function in Equation (\ref{RealnessGAN}). Thus, the loss for $D$ can be \gaoc{formulated} as:
\begin{equation}\label{loss_Denc}
\begin{aligned}
\mathcal{L}_{D_{enc}^{U}} =& -\mathbb{E}_{\boldsymbol{x} \sim p_{\text {data }}} \big[\mathcal{D}_{\mathrm{KL}}(\mathcal{R}_{1}(\mathbf{v}) \| D_{enc}^{U}(\boldsymbol{x}))+\log (D_{enc}^{U}(\boldsymbol{x}))\big] \\
&-\mathbb{E}_{\boldsymbol{x} \sim p_{\text {g }}}[\mathcal{D}_{\mathrm{KL}}(\mathcal{R}_{0}(\mathbf{v}) \| D_{enc}^{U}(f_{\text {G}}\left(\mathbf{x}\right))) \\
&+\log (1-D_{enc}^{U}(f_{\text {G}}\left(\mathbf{x}\right)))],
\end{aligned}
\end{equation}
where $D_{enc}^{U}$ \gaoc{is} the encoder of the discriminator (i.e., $D$ in Equation (\ref{RealnessGAN})). $f_{\text {G}}$ represents the generator $G$. Correspondingly, $G$ aims to \gaob{minimize} the value function in Equation (\ref{RealnessGAN}). Thus, the adversarial loss for $G$ becomes:

\begin{equation} \label{adv_loss1}
\begin{aligned}
\mathcal{L}_{\mathrm{adv1}}=&-\mathbb{E}_{\boldsymbol{x} \sim p_{g}}[\mathcal{D}_{\mathrm{KL}}(\mathcal{R}_{0}(\mathbf{v}) \| D_{enc}^{U}(f_{\text {G}}\left(\mathbf{x}\right)))\\
&+\log (1-D_{enc}^{U}(f_{\text {G}}(\mathbf{x})))],
\end{aligned}
\end{equation}
The \gaob{optimization} of Equation (\ref{RealnessGAN}) requires that \gaob{HP-ALF} reaches the Nash equilibrium. This further leads to the optimality of the generator and the discriminator. Specifically, \gaob{Theorem 1} states that $D$ can reach its optimality for any given generator $G$. Then, in Theorem 2, $G$ can also reach its optimality when $D$ satisfies this optimality condition. The proofs for \gaob{Theorems 1 and 2} are \gaoc{presented} in Appendix \ref{Appendix2}.


\begin{theorem}\label{Theorem1}
When $G$ is fixed, for any outcome $\mathbf{v}$ and input sample $x$, the optimal discriminator $D$ satisfies
$$
D_{G}^{\star}(\boldsymbol{x}, v)=\frac{ p_{\text {data }}(\boldsymbol{x})}{p_{\text {data }}(\boldsymbol{x})+p_{g}(\boldsymbol{x})}+
\frac{\mathcal{\mathcal { R }}_{1}(v) p_{\text {data }}(\boldsymbol{x})+\mathcal{R}_{0}(v) p_{g}(\boldsymbol{x})}{p_{\text {data }}(\boldsymbol{x})+p_{g}(\boldsymbol{x})}
$$
\end{theorem}

\vspace{1em}
\begin{theorem}
When $D=D_{G}^{\star}$, and there exists an outcome $v \in \Omega$ such that $\mathcal{R}_{1}(v) \neq \mathcal{R}_{0}(v)$, the maximum of $V\left(G, D_{G}^{\star}\right)$ is achieved if and only if $p_{g}=p_{\text {data }}$
\end{theorem}

\subsection{Global and Local Coherent Discriminator}\label{sec:GLC Dis}
HP-ALF \gaob{utilizes} the decoder of the global and local coherent discriminator (\gaob{shown in Figure \ref{Figure25}}) to achieve the \gaob{regularization} term \gaor{$\sum_{i,j}\left\|\left[f_\text{D}\left(\mathbf{x}\right)\right]_{i,j}-[f_\text{D}\left(f_\text{G}\left(\mathbf{x}_{u}\right)\right)]_{i,j}\right\|$} in Equation (\ref{SolutionCS}). This decoder can improve the ability to reconstruct globally and locally coherent images in existing adversarial learning. This is because, in \gaoc{the} existing adversarial learning, these discriminators often focus either on the global structure or local details and thus provide insufficient information for the generator \cite{schonfeld2020u}. In contrast, HP-ALF \gaoc{builds} the global and local coherent discriminator to provide \gaoc{the} detailed per-pixel decision to the generator while maintaining the global coherence of reconstructed images.
The decoder of this discriminator in HP-ALF is constructed based on the term \gaor{$\sum_{i,j}\left\|\left[f_\text{D}\left(\mathbf{x}\right)\right]_{i,j}-[f_\text{D}\left(f_\text{G}\left(\mathbf{x}_{u}\right)\right)]_{i,j}\right\|$}. It outputs the classification on every pixel $(i,j)$ and then calculates the classification difference over all pixels between $x_{t}$ and $f_\text{G}\left(x_{u}\right)$. The \gaob{optimization} process of the decoder can be formulated in a loss function. Then, this loss function can construct an adversarial loss for the generator to receive the global and local feedback from the decoder.

\begin{figure}[t!]
\centering
	\includegraphics[width=0.49\textwidth]{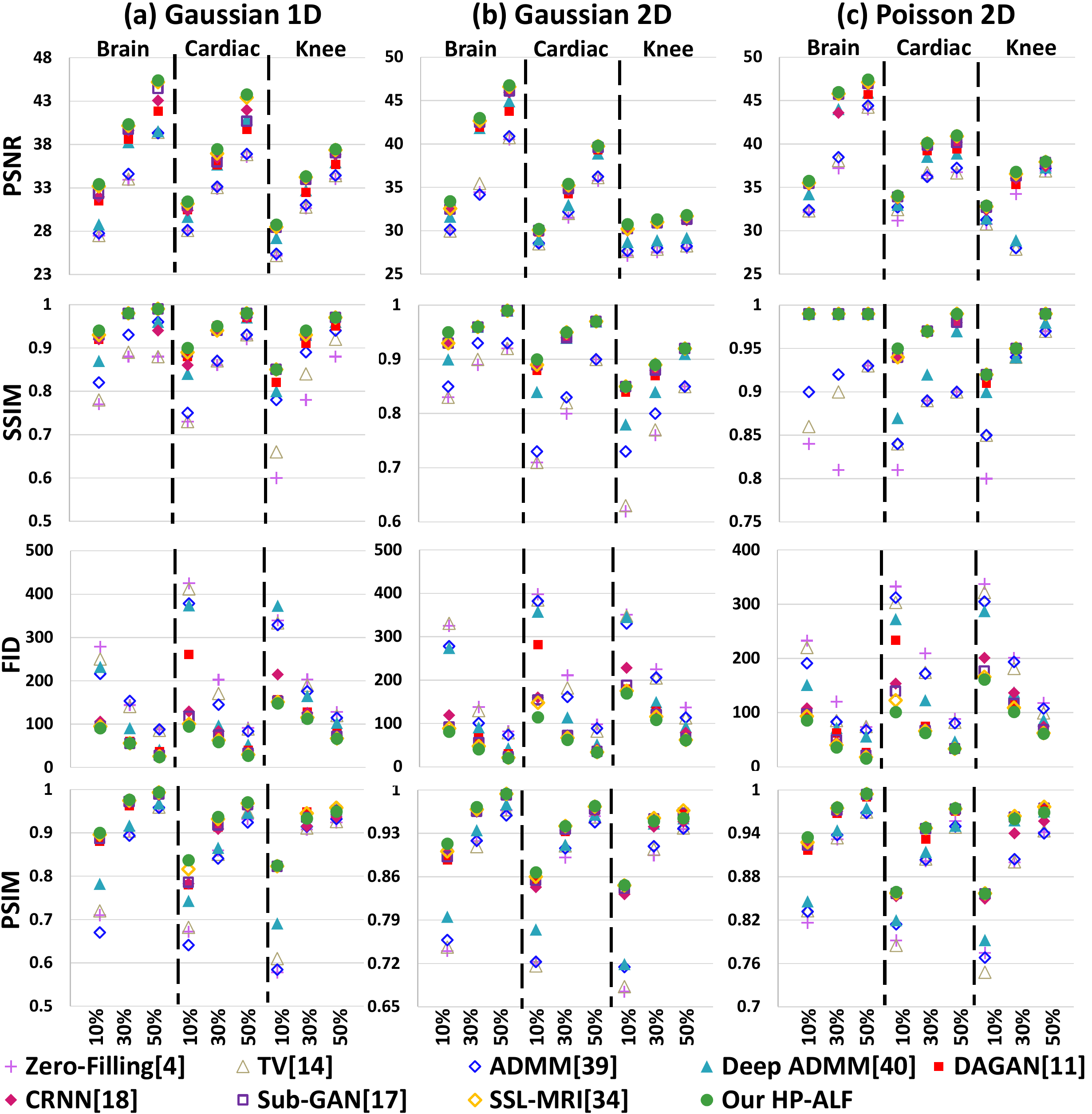}
\caption{Quantitative results (PSNR, SSIM, FID and PSIM) of the comparison study using different random undersampling masks (Gaussian 1D, Gaussian 2D and Poisson 2D). 10\%, 30\%, 50\% represent the percentage of the data sampling in the \textit{k}-space \gaob{data obtained} from original image data.}
\label{Table1}
\end{figure}

\begin{figure}[t]
\centering
	\includegraphics[width=0.49\textwidth]{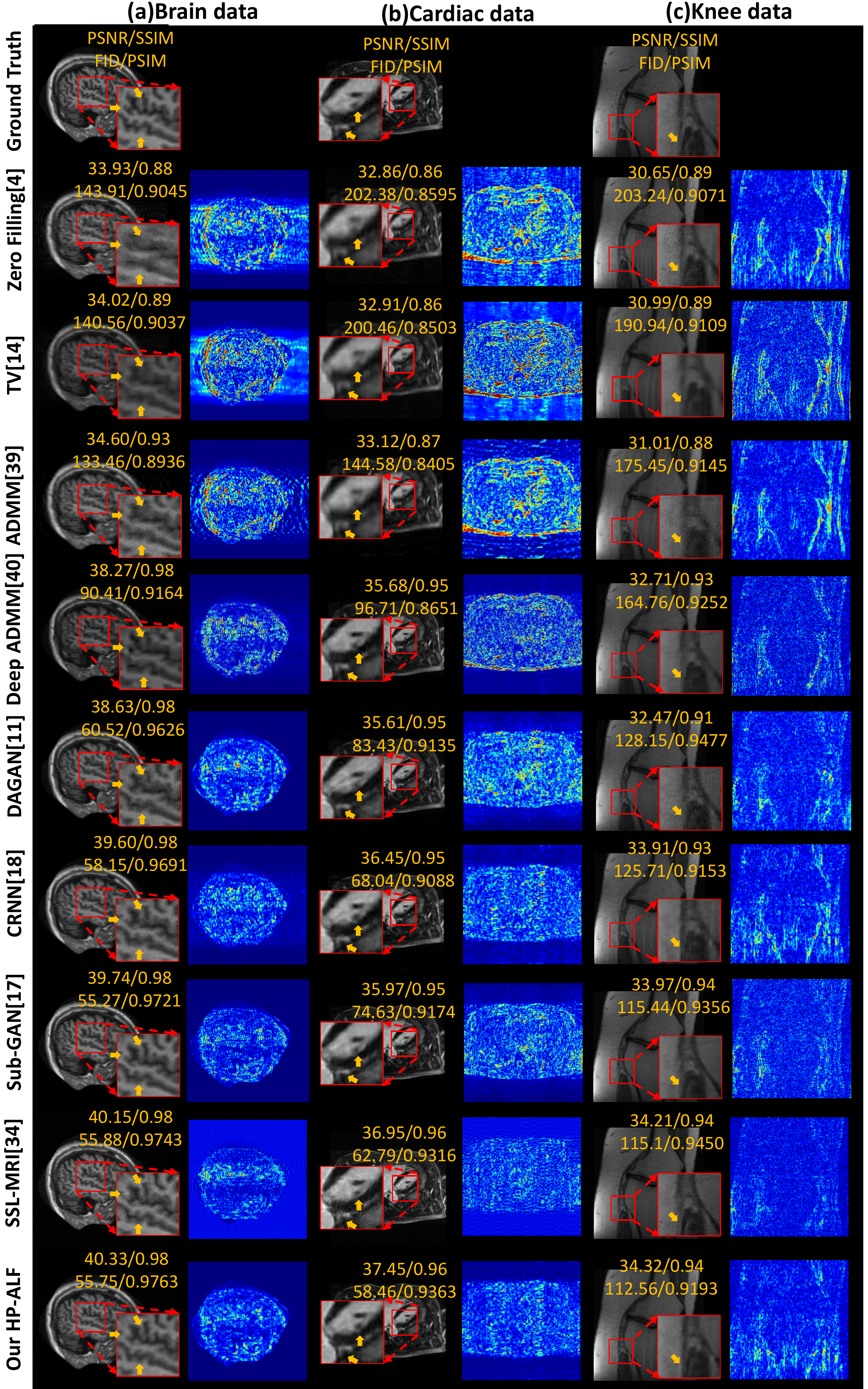}
\caption{Qualitative comparison with some representative methods using 30\% of the \textit{k}-space data and 1D Gaussian mask. Red boxes illustrate the enlarged view. The right panel images in different datasets illustrate the difference view. The numbers in the images are the PSNR, SSIM, FID and PSIM values.}
\label{Figure3}
\end{figure}

Specifically, the discriminator $D^{U}(x)$ \gaob{has} a \gaob{U-net-based} architecture including the \gaor{encoder} $D_{e n c}^{U}$ and the \gaor{decoder} $D_{dec}^{U}$.
\gaom{The output of encoder $D_{enc}^{U}$ includes two parts. The first part is the feature map at the last layer in the encoder. It is applied for the \gaob{multilevel} perspective discrimination. This enables the \gaob{encoder} $D_{enc}^{U}$ \gaob{to act} as \gaob{a} discriminator to classify the image \gaob{as} real or fake\gaob{,} as mentioned in \gaob{Section} \ref{sec:method RealnessGAN}. The second part includes the feature maps from the multiple levels in the encoder.
The input of the decoder $D_{dec}^{U}$ includes the two parts of the output of the encoder. The first part is fed into the decoder for \gaob{subsequent} upsampling. The second part of the encoder output \gaob{is} fed to the corresponding levels in the decoder by skip connection. It can perform the classification on a per-pixel basis, segmenting image $x$ into real and fake regions.}
Therefore, the discriminator loss can be formulated by taking the decisions from both $D_{e n c}^{U}$ and $D_{d e c}^{U}$:
$\mathcal{L}_{D^{U}}=\mathcal{L}_{D_{e n c}^{U}}+\mathcal{L}_{D_{d e c}^{U}}$,
where $D_{e n c}^{U}$ outputs the discrimination distribution score from the \gaob{multilevel} perspective discrimination and $D_{d e c}^{U}$ outputs the per-pixel decision.
The loss for the decoder $\mathcal{L}_{D_{d e c}^{U}}$ is formulated as the mean decision over all pixels:
\begin{equation}
\begin{aligned}
\mathcal{L}_{D_{d e c}^{U}}= &-\mathbb{E}_{\boldsymbol{x} \sim p_{\text {data }}} \big[\sum_{i, j} \log \left[D_{d e c}^{U}(x)\right]_{i, j}\big] \\
&-\mathbb{E}_{\boldsymbol{x} \sim p_{\text {g }}}\big[\sum_{i, j} \log \big(1-\left[D_{d e c}^{U}(G(x))\right]_{i, j}\big)\big],
\end{aligned}
\label{Unet_discriminator}
\end{equation}
where $\left[D_{d e c}^{U}(x)\right]_{i, j}$ and $\left[D_{d e c}^{U}(G(z))\right]_{i, j}$ are the local (per-pixel) \gaob{decisions} of the images at pixel $(i,j)$. HP-ALF encourages the discriminator to provide the detailed per-pixel decision to the generator. Therefore, the discriminator can \gaob{help} the generator maintain the global coherence of the reconstructed images by providing \gaob{global} image feedback. Correspondingly, the adversarial loss in Equation (\ref{Unet_discriminator}) becomes:

\begin{equation} \label{adv_loss2}
\begin{aligned}
\mathcal{L}_{\mathrm{adv} 2}=-\mathbb{E}_{\boldsymbol{x} \sim p_{g}}\big[\sum_{i, j} \log \left[D_{d e c}^{U}\left(f_{\mathrm{G}}(\mathbf{x})\right)\right]_{i, j}\big].
\end{aligned}
\end{equation}

\subsection{Context-aware Learning Generator} \label{sec:method description2}
\gaoc{The} context-aware learning generator $f_{\text {G}}$ in Equation (\ref{SolutionCS}) \gaoc{aims to} reconstruct the \gaob{artifacts}-free MR images from $\mathbf{x}_{u}$. As shown in Figure \ref{Figure_model}(c), the context-aware learning generator includes a \gaob{U-net-based} architecture $f_{\text {Unet}}$ and a context-aware learning block $f_{\text {CAL}}$. The generator is defined as \gaob{follows:}
\begin{equation}\begin{aligned}
\hat{\mathbf{x}}_{u} =f_\text{Unet}\left(\mathbf{x}_{u}\right), \mathbf{x}_{rec} &=f_\text{CAL}\left(\hat{\mathbf{x}}_{u}\right),
\end{aligned}\end{equation}
where $\mathbf{x}_{u}$ is the input for the generator. $\hat{\mathbf{x}}_{u}$ and $\mathbf{x}_{rec}$ are the outputs of $f_{\text {Unet}}$ and $f_{\text {CAL}}$, respectively. The U-net based architecture fully \gaob{utilizes} the 2D spatial information in each slice, but neglects the correlation between adjacent 2D slices. The insufficient prior information leads to the inaccurate reconstruction of \gaoc{the} fine details \cite{guo2020deep}. Therefore, the context-aware learning block exploits the 3D spatial feature from the input sequence of the MRI data. Specifically, this block \gaob{utilizes} the enhancement ConvLSTM (\gaob{i.e.,} Bi-ConvLSTM) to achieve the exploration of 3D semantic knowledge.
\gaob{The} LSTM unit contains a memory cell $\mathcal{C}_{t}$, an input gate $i_{t}$, a forget gate $f_{t}$, an output gate $o_{t}$, and an output state $\mathcal{H}_{t}$. However, ConvLSTM \gaob{replaces} LSTM by fully connected transformations with spatial local convolutions. ConvLSTM can be formulated as follows:
\begin{equation}
\begin{array}{l}
i_{t}=\sigma\left(\mathbf{W}_{x i} * \mathcal{X}_{t}+\mathbf{W}_{h i} * \mathcal{H}_{t-1}+\mathbf{W}_{c i} * \mathcal{C}_{t-1}+b_{i}\right) \\
f_{t}=\sigma\left(\mathbf{W}_{x f} * \mathcal{X}_{t}+\mathbf{W}_{h f} * \mathcal{H}_{t-1}+\mathbf{W}_{c f} * \mathcal{C}_{t-1}+b_{f}\right) \\
\mathcal{C}_{t}=f_{t} \circ \mathcal{C}_{t-1}+i_{t} \tanh \left(\mathbf{W}_{x c} * \mathcal{X}_{t}+\mathbf{W}_{h c} * \mathcal{H}_{t-1}+b_{c}\right) \\
o_{t}=\sigma\left(\mathbf{W}_{x o} * \mathcal{X}_{t}+\mathbf{W}_{h o} * \mathcal{H}_{t-1}+\mathbf{W}_{c o} \circ \mathcal{C}_{t}+b_{c}\right) \\
\mathcal{H}_{t}=o_{t} \circ \tanh \left(\mathcal{C}_{t}\right),
\end{array}
\end{equation}
where $\sigma$ and $\circ$ denote the convolution and Hadamard functions, respectively. $\mathcal{X}_{t}$ is the input tensor. Bi-ConvLSTM uses two ConvLSTMs to process the input data into two directions of \gaoc{the} forward and backward paths \gaob{and} then makes a decision for the current input by dealing with the data dependencies in both directions. The output of Bi-ConvLSTM can be calculated as:
\begin{equation}
\mathbf{Y}_{t}=\tanh (\mathbf{W}_{y}^{\overrightarrow{\mathcal{H}}} * \overrightarrow{\mathcal{H}}_{t}+\mathbf{W}_{y}^{\overleftarrow{\mathcal{H}}} \overleftarrow{\mathcal{H}}_{t}+b),
\end{equation}
where $\overrightarrow{\mathcal{H}}_{t}$ and $\overleftarrow{\mathcal{H}}_{t}$ denote the hidden state tensors for \gaoc{the forward state and the backward state}, respectively, $\mathbf{W}_{y}^{\overrightarrow{\mathcal{H}}}$ and $\mathbf{W}_{y}^{\overleftarrow{\mathcal{H}}}$ denote the weight parameters for $\overrightarrow{\mathcal{H}}_{t}$ and $\overleftarrow{\mathcal{H}}_{t}$, \gaob{respectively}. $b$ is the bias term, and $\mathbf{Y}_{t}$ indicates the final output considering bidirectional information. Through \gaob{the} Bi-ConvLSTM module, HP-ALF can learn \gaoc{the} fine details of MRI data slices.

\begin{figure}[t!]
\centering
	\includegraphics[width=0.49\textwidth]{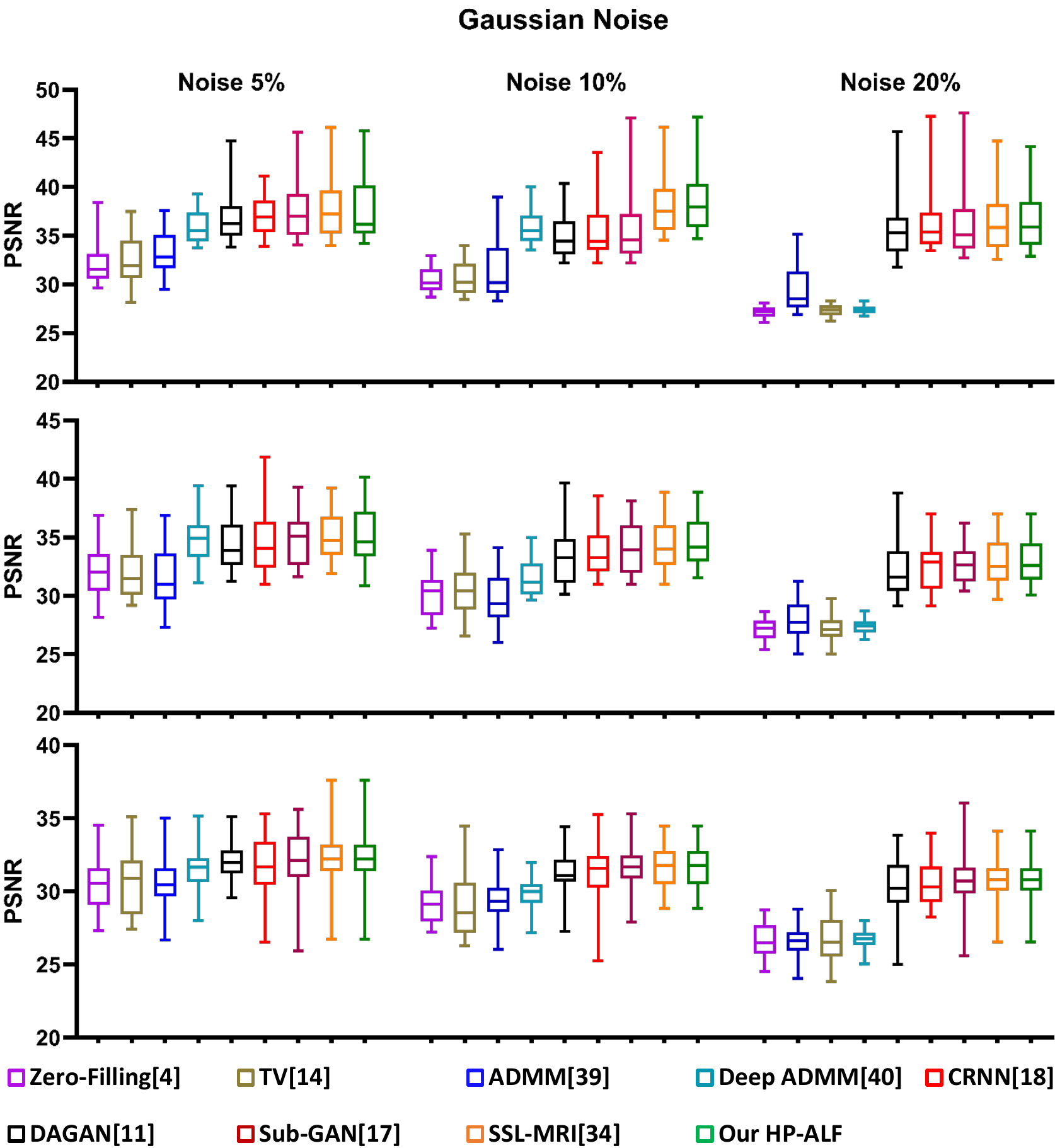}
\caption{The \gaob{comparison of reconstruction performance} using Gaussian noise in different noise levels. The comparison uses 1D Gaussian mask and 30\% percentage for data sampling. The upper, middle, and below panels show the results of the brain MRI dataset, the cardiac MRI dataset, and the knee MRI dataset respectively.}
\label{Figure4}
\end{figure}

\subsection{Loss Function and Implementation Details} \label{sec:method description6}
\gaoc{The loss function for Equation (\ref{SolutionCS}) consists of} a content loss and an adversarial loss. The content loss function is basically made up of a frequency-domain MSE loss and a perceptual VGG loss. The whole loss function can be formulated as
\begin{equation} \label{L_loss}
\mathcal{L}_{\mathrm{TOTAL}}=\alpha\mathcal{L}_{\mathrm{fMSE}}+\beta \mathcal{L}_{\mathrm{VGG}}+\mathcal{L}_{\mathrm{adv}},
\end{equation}
where $\alpha$ and $\beta$ \gaob{are hyperparameters}. {\color{black}{$\mathcal{L}_{\mathrm{adv}}$ represents the adversarial loss.}} $\mathcal{L}_{\mathrm{fMSE}}$ is the frequency-domain MSE loss and $\mathcal{L}_{\mathrm{VGG}}$ is the perceptual VGG loss.

The adversarial loss in Equation (\ref{L_loss}) includes two parts. First, $\mathcal{L}_{\mathrm{adv1}}$ is the adversarial loss based on the \gaob{optimization} process in Equation (\ref{RealnessGAN}). It can optimise $\int_{\mathbf{v}}\left(\left\|\mathbf{v}_{\mathbf{x}}-f_{\text {G}}\left(\mathbf{v}_{\mathbf{x}_{u}}\right)\right\|_{2}^{2}\right) d \mathbf{v}$ and $\|\mathbf{x}-f_{\text {G}}\left(\mathbf{x}_{u}\right)\|_{2}^{2}$ in Equation (\ref{SolutionCS}). Second, $\mathcal{L}_{\mathrm{adv2}}$ is the adversarial loss based on the loss function in Equation (\ref{Unet_discriminator}). It can \gaob{optimize the regularization} term $\mathcal{R}(\mathbf{x})$ in Equation (\ref{SolutionCS}). The whole adversarial loss function can be formulated as:
\begin{equation}
\begin{aligned}
&\mathcal{L}_{\mathrm{adv}}=-\mathbb{E}_{\boldsymbol{x} \sim p_{g}} \left[\mathcal{D}_{\mathrm{KL}}\left(\mathcal{R}_{0} \| D_{enc}^{U}(f_{\text {G}}\left(\mathbf{x}\right))\right)\right. \\
&+\log (1-D_{enc}^{U}(f_{\text {G}}\left(\mathbf{x}\right)))-\sum_{i, j} \log \left[D_{d e c}^{U}\left(f_{\mathrm{G}}(\mathbf{x})\right)\right]_{i, j}.
\end{aligned}
\end{equation}

The content loss can improve the perceptual quality of reconstruction. It includes a VGG loss and a frequency-domain MSE loss as constraints formulated as:
\begin{equation}\begin{aligned}
\mathcal{L}_{\mathrm{fMSE}} =\frac{1}{2}\left\|\mathrm{y}_{t}-\hat{\mathrm{y}}_{u}\right\|_{2}^{2}, \mathcal{L}_{\mathrm{VGG}}=\frac{1}{2}\left\|f_{\gaor{\mathrm{VGG}}}\left(\mathrm{x}_{t}\right)-f_{\gaor{\mathrm{VGG}}}\left(\hat{x}_{u}\right)\right\|_{2}^{2}.
\end{aligned}\end{equation}
$\mathcal{L}_{\mathrm{fMSE}}$ represents the difference between the reconstructed image and the original image in \gaob{the} frequency domain. Therefore, it can be \gaoc{applied} to achieve the data fidelity term $\lambda\left\|\mathbf{y}-\mathbf{F}_{u}\mathbf{x}\right\|_{2}^{2}$ in Equation (\ref{SolutionCS}). $\mathcal{L}_{\mathrm{VGG}}$ \gaoc{is an} additional regularisation term $\mathcal{R}(\mathbf{x})$ to constrain the solution space, where \gaor{$f_{\mathrm{VGG}}$} denotes VGG feature maps of VGG16.

HP-ALF consists of the context-aware learning generator and the global and local coherent discriminator.
The generator $G$ includes a \gaob{U-net-based} architecture and a context-aware learning block.
The \gaob{U-Net-based} architecture consists of eight convolutional layers (encoder layers) and \gaob{eight corresponding} deconvolutional layers (decoder layers).
The numbers of the filters are 64, 128, 256, 512, 512, 512, \gaob{and} 512 in \gaob{the} encoder layers and 1024, 1024, 1024, 1024, 512, 256, \gaob{and} 128 in \gaob{the} decoder layers. Each layer in $G$ uses a kernel \gaoc{with} size $k = 3$ with the batch \gaob{normalization} and leaky ReLU layers \gaob{behind it}.
The subsequent module of the \gaob{U-Net-based} architecture is the context-aware learning block. It has a Bi-ConvLSTM block, where the kernel size is $k = 3$ and the feature map channel inside is 32.
$G$ also applies refinement learning to connect layers between the input and the output of the context-aware learning block. The architecture of $D$ is similar to the \gaob{U-net-based} architecture of $G$.
In addition, $D$ cascades three dense convolutional layers after the encoder layer, and the sigmoid activation function outputs the classification results.

HP-ALF uses 5 consecutive 2D slices from 3D data as the input sequence. It also \gaob{normalizes} the intensities of all 2D slices into the range between -1 and 1 \cite{yang2017dagan}. These adjacent slices are fed into the network in chip orders. Then, the output of the U-net is reshaped \gaob{into the input} of the context-aware learning block.
\gaom{The discriminator uses the original or generated image as input with \gaob{a size of} 256$\times$256.
In the discriminator, the encoder represents and downsamples the input image to a feature map (channels=10, columns=4, rows=4).
This feature map is then transformed to a 10$\times$1 feature vector \gaob{by global} sum pooling. The decoder applies the feature maps as input and upsamples it in each level until it reaches the original image size \gaob{of} 256$\times$256.}

\begin{figure}[t!]
\centering
	\includegraphics[width=0.49\textwidth]{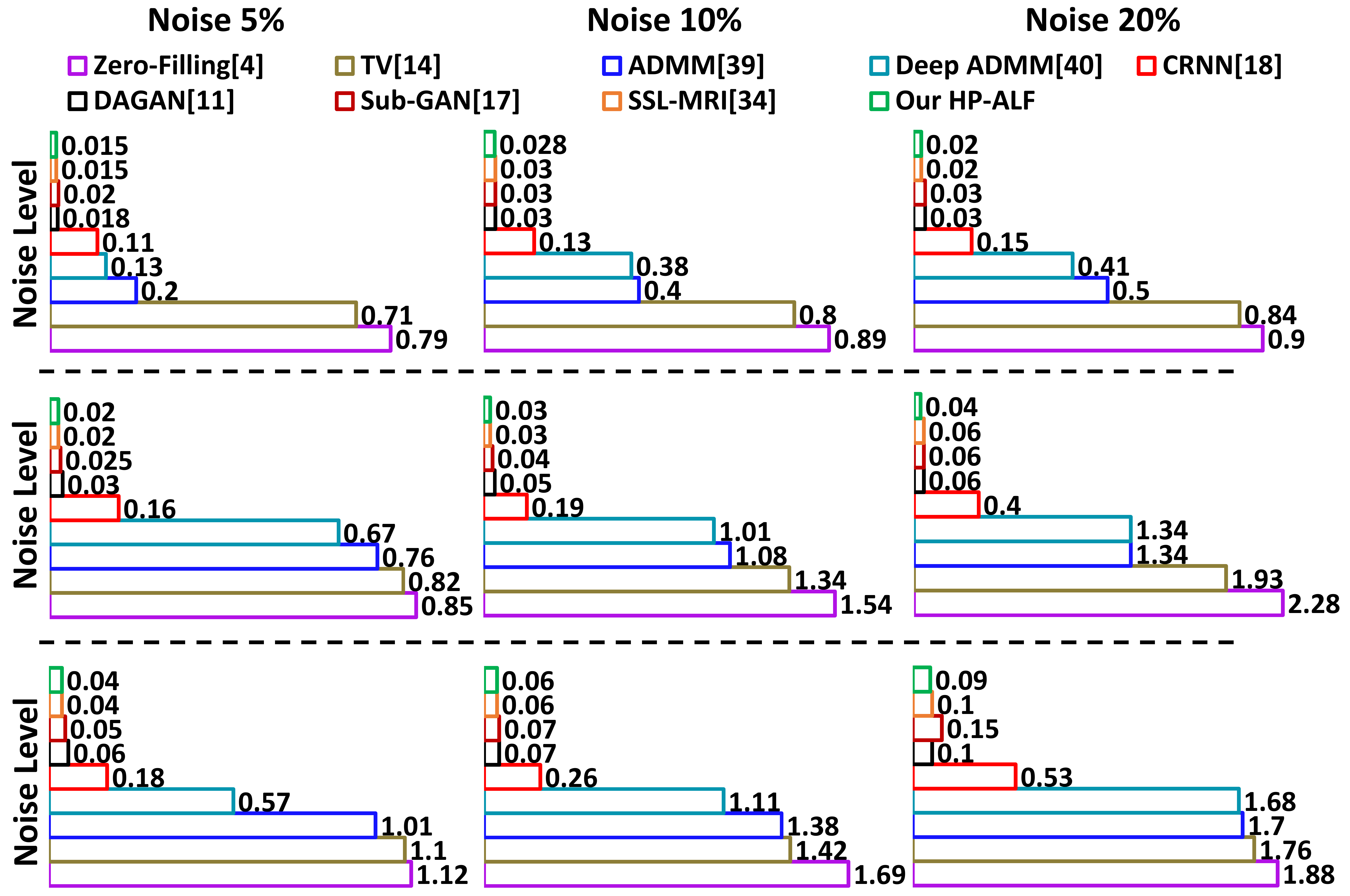}
\caption{\gaob{The comparison of noise reduction performance} using Gaussian noise in different noise levels \gaoc{by the residual noise level}. The comparison uses 1D Gaussian mask and 30\% percentage for data sampling. The upper, middle, and lower panels show the results of the brain MRI dataset, the cardiac dataset, and the knee dataset, respectively. The estimated value of the noise level is proportional to the residual noise.}
\label{Figure_NoiseLevel}
\end{figure}

\section{Experiment and Results}

\subsection{Data Collection}\label{sec:Data description}

The experiments are carried out on three MRI datasets. (1) Brain MRI dataset: This is the MICCAI 2013 grand challenge dataset\footnote{\url{http://masiweb.vuse.vanderbilt.edu/workshop2013/index.php/}}. It contains 150 3D patient data with 42750 slices. \gaoc{Each patient data} includes about 285 slices with 256$\times$256 pixels. (2) Cardiac MRI dataset: This is the 2018 Atrial Segmentation Challenge dataset\footnote{\url{https://atriaseg2018.cardiacatlas.org/}}. It contains 100 3D LGE MRI patient data with 5920 slices. \gaoc{Each patient data} includes about 60 slices with 256$\times$256 pixels. A whole-body MRI scanner is used for this dataset. The image acquisition resolution is 0.625mm$^2$. (3) Knee MRI dataset: This is the FastMRI dataset\footnote{\url{https://fastmri.org/dataset/}}. It contains 96 3D patient data with 3270 slices. \gaoc{Each patient data} includes about 35 slices with 256$\times$256 pixels. Each dataset is divided \gaob{into training} data (70\%), validation data (20\%) and test data (10\%). \gaoc{In} all 3D data, we exclude the slices at the edge, where the number \gaob{of void} pixels is greater than 90\%. For the three datasets, the DICOM data are collected, rather than the raw \textit{k}-space data. Not using the raw \textit{k}-space data results in the inequality of image quality between the ground truth image and the fully sampled raw data. This is because the raw \textit{k}-space data directly \gaob{correspond} to the originally measured raw data \cite{zbontar2018fastmri,sandino2020compressed}. However, the use of DICOM data considers the reproducibility and generality of the reconstruction method. This is \gaob{because clinical centers} usually save the image data \cite{sandino2020compressed}.

\begin{table}[]
\caption{Reconstruction time and Parameters of the comparison study. ``Num of Param'' means the numbers of the parameters in the compared models. \gaor{``$\pm$"} means standard deviation.}
\centering
\resizebox{0.35\textwidth}{!}{\begin{tabular}{l|c|c}
\hline
Methods & \begin{tabular}[c]{@{}c@{}}Testing Time\\ CPU(sec)/GPU(ms)\end{tabular} & Num of Param \\ \hline
Zero-Filling \cite{lustig2008compressed} & 0,002$\pm$0.003/ -   & -       \\
TV \cite{block2007undersampled}          & 10.5$\pm$1.2/ -      & -       \\
ADMM \cite{yang2010fast}        & 10.2$\pm$2.3/ -      & -       \\
Deep ADMM \cite{sun2016deep}   & 3.2$\pm$0.2/ -       & 397.19K \\
DAGAN \cite{yang2017dagan}       & 0.2$\pm$0.1/5.4$\pm$0.1 & 564.76M \\
CRNN \cite{qin2018convolutional}        & 0.2$\pm$0.1/6.3$\pm$0.3 & 1.14M   \\
Sub-GAN \cite{shaul2020subsampled}        & 0.2$\pm$0.1/5.9$\pm$0.4 & 566.7M   \\
SSL-MRI \cite{hu2021self}        & 0.2$\pm$0.1/4.2$\pm$0.5 & 2.68M   \\
Our HP-ALF       & 0.2$\pm$0.1/5.7$\pm$0.1 & 217.9M \\ \hline
\end{tabular}}
\label{Table_TimeandParam}
\end{table}

\gaob{To} reduce the extra computational burden, the strategy in \cite{yang2017dagan} is applied to handle the complex-valued data. Specifically, the real-valued information can be embedded into the complex space using an operator $\operatorname{Re}^{*}: \mathbb{R}^{N} \mapsto \mathbb{C}^{N}$ such that $\mathrm{Re}^{*}(\mathrm{x})=\mathrm{x}+0 i$, and therefore the MRI forward operator can be expressed as $\mathrm{F}: \mathbb{R}^{N} \stackrel{\mathrm{Re}^{*}}{\mapsto} \mathbb{C}^{N} \mapsto \mathcal{F} \mathbb{C}^{N} \mapsto \mathcal{U} \mathbb{C}^{M}$, \gaob{where} $\mathrm{F}_{u}$ combines the Fourier transform $\mathcal{F}$ and random \gaob{undersampling operators $\mathcal{U}$}.

The experiments \gaob{utilize} the single-coil MRI data for training, \gaoc{although} most existing methods \gaob{utilize} the multi-coil MRI data. This is because HP-ALF \gaob{utilizes} the zero-filling images that come from the preprocessing of raw multi-coil or single-coil data, while existing methods \gaob{utilize} the multi-coil MRI data to extract the coil sensitivity \cite{zbontar2018fastmri}. Therefore, it does not affect HP-ALF whether the data input is \gaob{single-coil} or multi-coil in the reconstruction process. Moreover, the single-coil MRI image obtained by data preprocessing is of lower quality at the same acceleration factor\cite{zbontar2018fastmri}. Therefore, it is more challenging for HP-ALF to use the single-coil MRI data as input compared with the multi-coil MRI data.
In addition, HP-ALF, \gaob{similar to} other GAN-based methods, can combine the parallel imaging strategy \gaob{and transfer} learning for \gaob{multichannel imaging} \cite{lv2021pic,lv2021transfer}.  The extension to multi-coil MRI data will be considered in the future studies.

\subsection{Evaluation Metrics and Training Details}
The evaluation metrics include the \gaob{peak signal-to-noise ratio} (PSNR), the structural similarity index (SSIM), the Fr$\acute{e}$chet inception distance (FID) and \gaob{the} perceptual similarity \gaob{measure (PSIM)}. PSNR evaluates the perceptual quality of reconstruction \cite{blau2018perception}. SSIM measures the perceptual similarity of images \cite{wang2004image}. FID is a similarity measure between two datasets that correlates well with \gaob{human judgments} of visual quality. It evaluates the similarity between the set of generated images and the corresponding fully sampled images \cite{heusel2017gans}. PSIM is a perceptual image quality assessment (IQA) metric based on the human visual system. It evaluates the similarity of local details between the input original and distorted images \cite{gu2017fast}.

HP-ALF uses the Adam \gaob{optimizer} with a batch of \gaoc{four} subjects per step and \gaob{an} initial learning rate of 0.0003 during the training process. To balance the weights of different losses in Equation (\ref{L_loss}) into similar scales, $\alpha$ is set \gaob{to} 15 and $\beta$ is set \gaob{to} 0.1 according to \cite{yang2017dagan}. The learning rate is halved every 5 epochs. \gaob{Early stopping} is used when \gaoc{the} validation loss stops decreasing for 50 epochs.

\begin{figure}
\centering
	\includegraphics[width=0.49\textwidth]{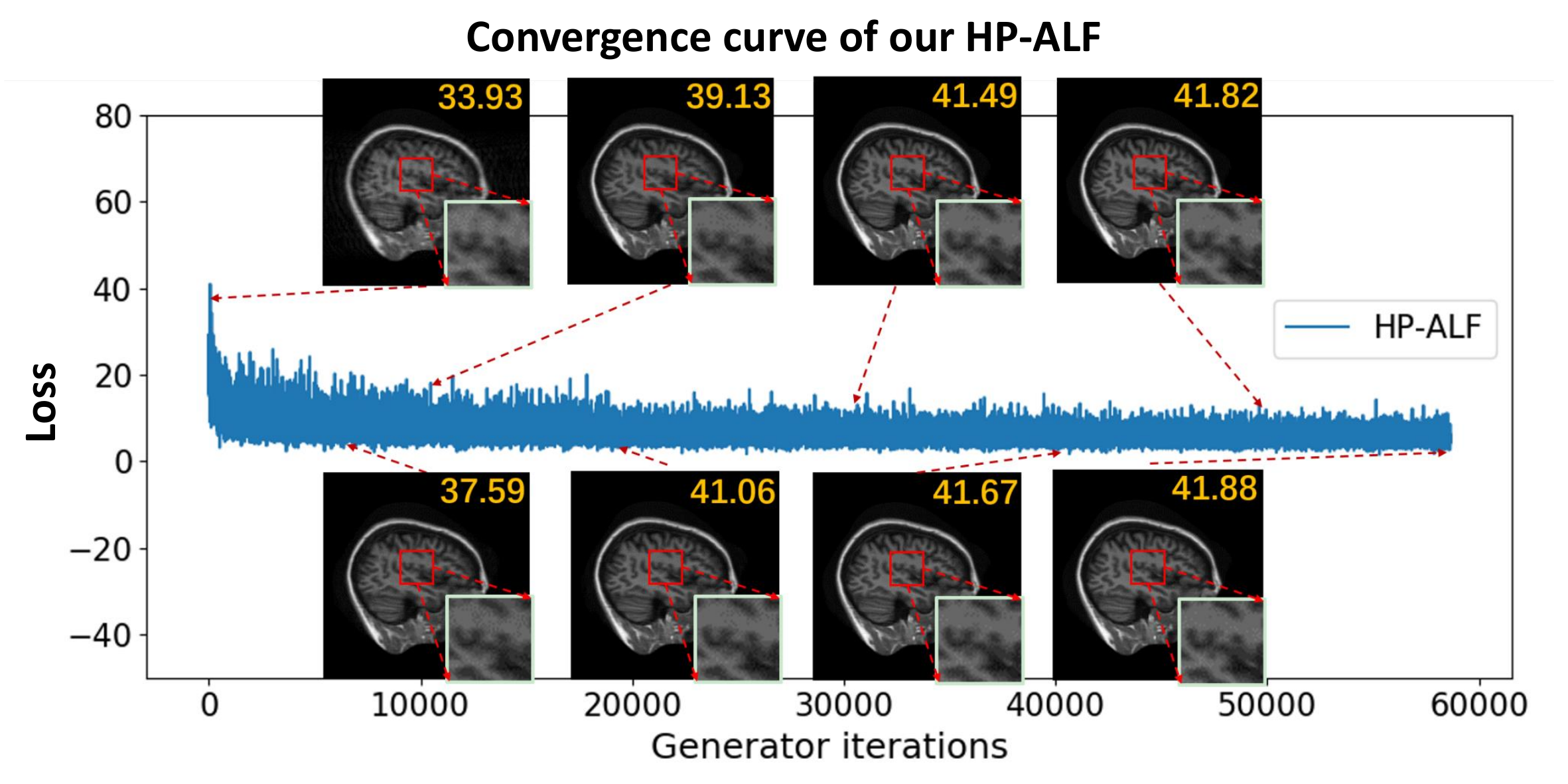}
\caption{Convergence validation for multi-level perspective discrimination using training curves and samples. It can be seen a clear correlation between lower error and better sample perception quality. The numbers in the images are the PSNR values.}
\label{Figure_RealnessSamples}
\end{figure}

\subsection{Results}

\textbf{Comparison with CS-MRI Methods.} We compare HP-ALF with seven CS-MRI methods, \gaob{including conventional} methods and deep-learning-based methods. The conventional methods include the \gaob{total variation} (TV) \cite{block2007undersampled} and ADMM \cite{yang2010fast}. The deep-learning-based methods include Deep ADMM \cite{sun2016deep}, DAGAN \cite{yang2017dagan}, CRNN \cite{qin2018convolutional}, Sub-GAN \cite{shaul2020subsampled} and SSL-MRI \cite{hu2021self}. All comparison methods and HP-ALF {\color{black}{use}} the baseline zero-filling reconstruction for \gaob{initialization} to achieve the fair comparison. In addition, all comparison methods use the default parameters recommended in their papers.

\gaoc{Figure \ref{Table1} shows the results of the method comparison \gaob{under} different undersampling conditions}. The image data \gaob{are first} transformed to the \textit{k}-space data. Then the \textit{k}-space data \gaob{are} undersampled using three masks: 1D Gaussian (G1D), 2D Gaussian (G2D), and 2D Poisson disc (P2D). Each mask retains $10\%, 30\%$, \gaob{and} $50\%$ of the data to achieve the 10$\times$, 3.3$\times$, and 2$\times$ acceleration, respectively. The results show that HP-ALF achieves the best scores in PSNR, SSIM and FID,
\gaob{especially} for the 1D Gaussian disc mask at 10$\times$ speed-up. \gaob{The values} of PSNR, SSIM, FID and PSIM \gaob{obtained} by HP-ALF are 32.42, 0.94, 90.79, \gaob{and} 0.90, \gaob{respectively}, in brain data; 33.41, 0.95, 80.72, \gaob{and} 0.91, \gaob{respectively}, in cardiac data; and 35.75, 0.99, 77.54, and 0.93, \gaob{respectively}, in knee data. \gaob{These values} are better than the other methods. However, DAGAN, \gaob{deep} ADMM and SSL-MRI can achieve higher PSIM values in knee data as 0.94, 0.93, \gaob{and} 0.94, \gaob{respectively}, at 3.3$\times$ and 0.95, 0.94, 0.95, \gaob{respectively}, at 2$\times$ speed-up.

\begin{table}[]
\caption{Quantitative results (PSNR and SSIM) of the ablation study using 1D Gaussian mask. 10\%, 30\%, 50\% represents the percentages of the data sampling in the \textit{k}-space data obtained from original image data. ``Num of Param'' means the numbers of the parameters. The abbreviations of the comparative methods are explained in section \ref{Experiment:Ablation Study}. \gaor{``$\pm$"} means standard deviation.}
\resizebox{0.48\textwidth}{!}{\renewcommand\arraystretch{1.05}{
\setlength{\tabcolsep}{2mm}{\begin{tabular}{cccccc}
\hline
Task                           & Mask                  & Methods      & PSNR                    & SSIM                    & Num of Param       \\
\hline
\multirow{15}{*}{\rotatebox{90}{Brain data}}   & \multirow{5}{*}{10\%} & with TAL & 28.01$\pm$2.31          & 0.87$\pm$0.02           & 217.9m           \\
                               &                       & without MPD  & 31.86$\pm$3.09          & 0.93$\pm$0.02           & 217.9m           \\
                               &                       & without GLC  & 32.04$\pm$3.35          & 0.94$\pm$0.02           & \textbf{217.6m}  \\
                               &                       & without CAL  & 32.19$\pm$2.98          & 0.94$\pm$0.02           & 217.9m           \\
                               &                       & HP-ALF       & \textbf{32.42$\pm$3.28} & \textbf{0.94$\pm$0.02}  & 217.9m           \\
\cline{2-6}
                               & \multirow{5}{*}{30\%} & with TAL & 35.01$\pm$4.48          & 0.95$\pm$0.02           & 217.9m           \\
                               &                       & without MPD  & 39.18$\pm$3.40          & 0.98$\pm$0.01           & 217.9m           \\
                               &                       & without GLC  & 39.83$\pm$3.13          & 0.98$\pm$0.01           & \textbf{217.6m}  \\
                               &                       & without CAL  & 39.36$\pm$3.32          & 0.98$\pm$0.01           & 217.9m           \\
                               &                       & HP-ALF       & \textbf{40.33$\pm$3.37} & \textbf{0.98$\pm$0.01}  & 217.9m           \\
\cline{2-6}
                               & \multirow{5}{*}{50\%} & with TAL & 39.67$\pm$2.85          & 0.96$\pm$0.01           & 217.9m           \\
                               &                       & without MPD  & 43.36$\pm$2.51          & 0.99$\pm$0.001          & 217.9m           \\
                               &                       & without GLC  & 44.68$\pm$2.54          & 0.99$\pm$0.001          & \textbf{217.6m}  \\
                               &                       & without CAL  & 44.05$\pm$2.10          & 0.99$\pm$0.001          & 217.9m           \\
                               &                       & HP-ALF       & \textbf{45.38$\pm$2.56} & \textbf{0.99$\pm$0.001} & 217.9m           \\
\hline
\multirow{15}{*}{\rotatebox{90}{Cardiac data}} & \multirow{5}{*}{10\%} & with TAL & 29.47$\pm$3.16          & 0.87$\pm$0.03           & 217.9m           \\
                               &                       & without MPD  & 30.21$\pm$2.78          & 0.86$\pm$0.02           & 217.9m           \\
                               &                       & without GLC  & 31.19$\pm$3.27          & 0.90$\pm$0.02           & \textbf{217.6m}  \\
                               &                       & without CAL  & 30.86$\pm$2.99          & 0.89$\pm$0.02           & 217.9m           \\
                               &                       & HP-ALF       & \textbf{31.41$\pm$3.18} & \textbf{0.90$\pm$0.02}  & 217.9m           \\
\cline{2-6}
                               & \multirow{5}{*}{30\%} & with TAL & 35.43$\pm$4.12          & 0.95$\pm$0.02           & 217.9m           \\
                               &                       & without MPD  & 36.18$\pm$2.56          & 0.95$\pm$0.01           & 217.9m           \\
                               &                       & without GLC  & 37.30$\pm$2.43          & 0.96$\pm$0.01           & \textbf{217.6m}  \\
                               &                       & without CAL  & 37.01$\pm$2.76          & 0.96$\pm$0.01           & 217.9m           \\
                               &                       & HP-ALF       & \textbf{37.45$\pm$1.99} & \textbf{0.96$\pm$0.01}  & 217.9m           \\
\cline{2-6}
                               & \multirow{5}{*}{50\%} & with TAL & 39.37$\pm$2.11          & 0.92$\pm$0.01           & 217.9m           \\
                               &                       & without MPD  & 41.98$\pm$2.84          & 0.97$\pm$0.01           & 217.9m           \\
                               &                       & without GLC  & 42.60$\pm$2.50          & 0.97$\pm$0.01           & \textbf{217.6m}  \\
                               &                       & without CAL  & 42.46$\pm$2.74          & 0.97$\pm$0.01           & 217.9m           \\
                               &                       & HP-ALF       & \textbf{42.77$\pm$2.34} & \textbf{0.97$\pm$0.01}  & 217.9m           \\
\hline
\multirow{15}{*}{\rotatebox{90}{Knee data}}    & \multirow{5}{*}{10\%} & with TAL & 27.45$\pm$2.31          & 0.80$\pm$0.06           & 217.9m           \\
                               &                       & without MPD  & 27.05$\pm$3.39          & 0.83$\pm$0.06           & 217.9m           \\
                               &                       & without GLC  & 28.07$\pm$3.41          & 0.85$\pm$0.06           & \textbf{217.6m}  \\
                               &                       & without CAL  & 27.49$\pm$2.76          & 0.85$\pm$0.06           & 217.9m           \\
                               &                       & HP-ALF       & \textbf{28.75$\pm$2.22} & \textbf{0.85$\pm$0.06}  & 217.9m           \\
\cline{2-6}
                               & \multirow{5}{*}{30\%} & with TAL & 31.99$\pm$1.97          & 0.91$\pm$0.03           & 217.9m           \\
                               &                       & without MPD  & 32.98$\pm$2.04          & 0.93$\pm$0.03           & 217.9m           \\
                               &                       & without GLC  & 34.02$\pm$2.13          & 0.94$\pm$0.03           & \textbf{217.6m}  \\
                               &                       & without CAL  & 33.79$\pm$1.79          & 0.93$\pm$0.03           & 217.9m           \\
                               &                       & HP-ALF       & \textbf{34.32$\pm$1.98} & \textbf{0.94$\pm$0.03}  & 217.9m           \\
\cline{2-6}
                               & \multirow{5}{*}{50\%} & with TAL & 34.86$\pm$2.71          & 0.95$\pm$0.01           & 217.9m           \\
                               &                       & without MPD  & 37.01$\pm$2.51          & 0.97$\pm$0.01           & 217.9m           \\
                               &                       & without GLC  & 37.28$\pm$2.54          & 0.97$\pm$0.01           & \textbf{217.6m}  \\
                               &                       & without CAL  & 37.19$\pm$2.10          & 0.97$\pm$0.01           & 217.9m           \\
                               &                       & HP-ALF       & \textbf{37.44$\pm$2.18} & \textbf{0.97$\pm$0.01}  & 217.9m           \\
\hline
\end{tabular}}}}
\label{Table_ha}
\end{table}

\gaoc{Figure \ref{Figure3} shows the visualization of \gaob{the} method comparison for the sample reconstructed images}. The results show that HP-ALF returns the sharpest images with fine details in brain data and cardiac data, as apparent from the magnified regions. Although CRNN and DAGAN can also return \gaoc{the} sharp images, the \gaoc{reconstruction} of the fine tissue structure \gaob{is} less detailed than \gaob{that} reconstructed by HP-ALF. \gaoc{Then}, Sub-GAN and SSL-MRI reconstruct the textures with less noise, but with unsatisfactory performance \gaob{in reconstructing} the fine details. For example, the areas within the block in Figure \ref{Figure3} show the unrealistic texture details in Sub-GAN and SSL-MRI. Moreover, ADMM and zero-filling reconstruction \gaob{have difficulty inhibiting} the remaining noise-like textures. However, Deep ADMM, DAGAN, Sub-GAN and SSL-MRI can achieve smaller error maps and retain fewer aliasing \gaob{artifacts}. Furthermore, \gaoc{these methods have better \gaob{PSIM performance} in knee data as the values are 0.9252, 0.9477, 0.9356 and 0.9450, \gaob{respectively}}. This is because the evaluation of \gaoc{the} local and detailed information by PSNR, SSIM and FID \gaoc{are} affected by global similarity \gaob{from} the perspective of \gaoc{the} single-scale structure.
\gaoc{Then this evaluation may be also affected by the \gaob{nonsalient} area (e.g., background) with fewer aliasing \gaob{artifacts}, when the remaining aliasing \gaob{artifacts} mainly appear in the bone areas of the local image}
In addition, the \gaoc{Knee dataset} contains more random noise. The random noise and texture details are indistinguishable with respect to other datasets. Thus, HP-ALF may perceive random noise as the texture detail of the image during learning.

Table \ref{Table_TimeandParam} presents the number of parameters and the reconstruction time of \gaoc{all methods}. \gaoc{The results show} that HP-ALF has the third largest number of parameters and the third fastest speed \gaob{among} all methods. Although Deep ADMM and CRNN have \gaob{fewer} numbers of \gaoc{the} parameters \gaob{than HP-ALF}, they \gaoc{do not} complete the reconstruction at the same time. Similarly, although DAGAN is faster than HP-ALF, it has \gaob{more parameters}. SSL-MRI can achieve the lowest number of parameters and the fastest speed \gaoc{in} all methods, but its performance does not reach \gaob{that of} HP-ALF.

\begin{figure}
\centering
	\includegraphics[width=0.49\textwidth]{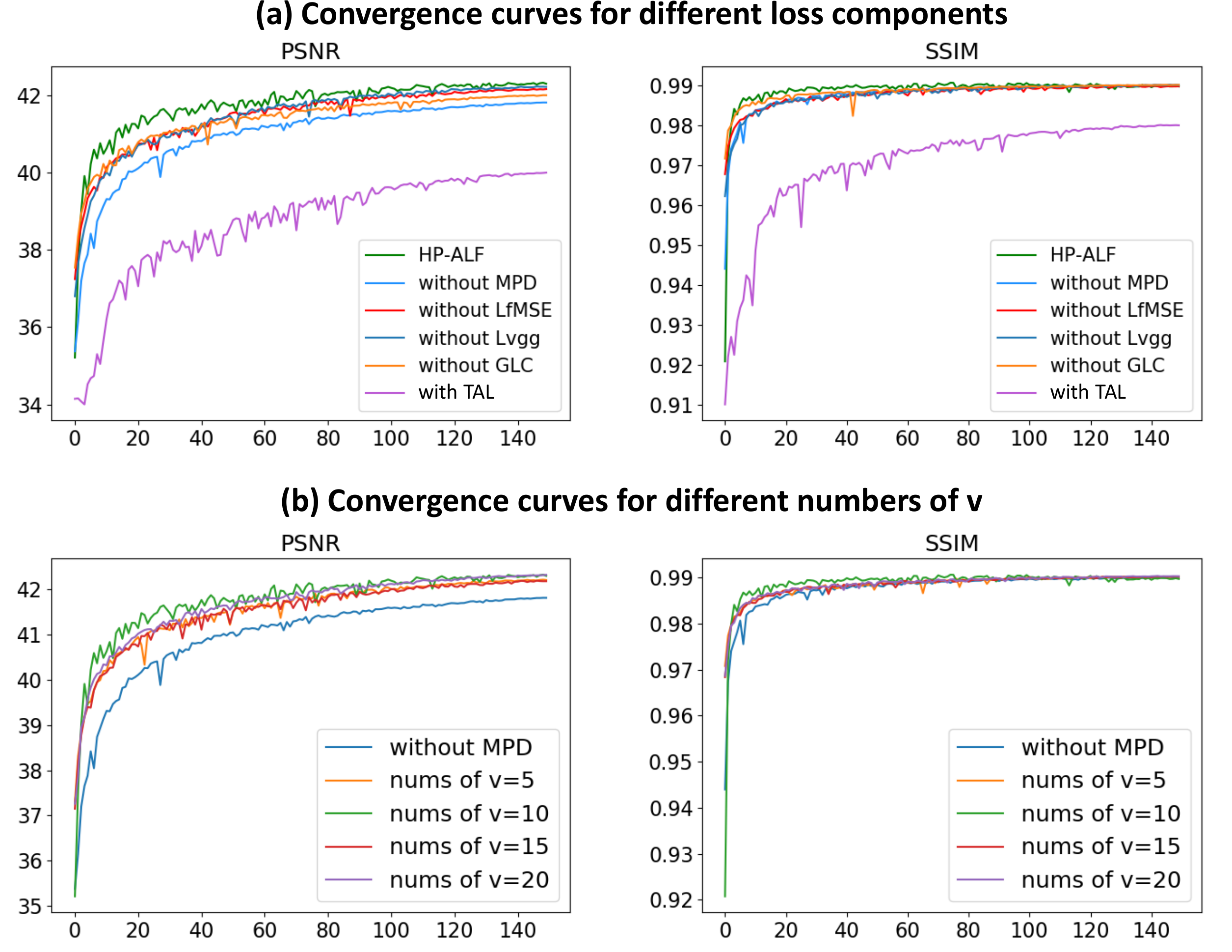}
\caption{(a) Convergence curves for different loss components in Equation (\ref{L_loss}). It can be seen the image quality \gaob{for} all loss functions increase quickly and converge after several epochs. (b) The comparison of the different number of $\mathbf{v}$ in $p_{\text {perspective }}$. When the number of $\mathbf{v}$ exceeds 10 (i.e., the default setting), the image quality does not improve significantly.}
\label{Figure_loss}
\end{figure}

\vspace{1.5mm}
\textbf{Noise Suppression Comparison.}
\gaoc{It compares the noise suppression performance of HP-ALF with \gaob{that of} the comparative methods.}
\gaoc{Gaussian noise is used because it} is suitable to simulate the natural noise of MRI. The natural noise of MRI mainly comes from the thermal noise of the scanned object. Gaussian noise is commonly used for \gaob{simulating thermal} noise in
\textit{k}-space \cite{sagheer2020review,Aja-Fernandez2016,henkelman1985measurement}. Moreover, Gaussian noise is also widely used in existing CS-MRI methods \cite{you2021coast,liu2020deep,deora2020structure}. These methods motivate us to insert Gaussian noise in \textit{k}-space for the noise suppression experiment. Specifically, \gaoc{this experiment} adds \gaob{additive white} Gaussian noise to the \textit{k}-space before applying the undersampling during training and testing. For the comparative analysis, \gaob{the} Gaussian noise is varied from 5\% to 20\%. \gaob{In addition}, the comparison study uses \gaob{a} 1D Gaussian mask and is performed on 30\% data sampling.
Additionally, the noise level estimation method in \cite{liu2013single} is applied to \gaoc{evaluate the residual noise \gaob{after noise} suppression}.

\gaoc{Figure \ref{Figure4} shows the noise suppression performance of HP-ALF at different noise levels by PSNR}. \gaoc{The PSNR of HP-ALF is a higher mean value than \gaob{those of the} other methods}. In particular, HP-ALF achieves the best PSNR values in all the datasets at the highest Gaussian noise level (i.e., noise 20\%) \gaob{with} 36.28 for \gaob{the} brain data, 32.91 for \gaob{the} cardiac data and 30.78 for \gaob{the} knee data.
\gaoc{\gaob{Figure \ref{Figure_NoiseLevel}} shows the noise suppression performance of HP-ALF at different noise levels by the residual noise level}. Especially for the highest Gaussian noise level (i.e., noise 20\%), HP-ALF achieves the lowest value of the residual noise levels in all the datasets \gaob{with} 0.02 for \gaob{the} brain data, 0.04 for \gaob{the} cardiac data, and 0.09 for \gaob{the} knee data. These results indicate that HP-ALF can effectively reduce \gaoc{the} residual noise in the reconstructed images.

\vspace{1.5mm}
\textbf{Performance of Different Components in HP-ALF.}\label{Experiment:Ablation Study}
The ablation study \gaoc{evaluates} how \gaob{changes in} the main components of HP-ALF \gaoc{affects} its performance. (1) without MPD: the method without \gaoc{the} \gaob{multilevel} perspective discrimination; (2) without CAL: the method without \gaoc{the}context-aware learning block; (3) without GLC: the method without \gaoc{the} global and local coherent discriminator; (4) HP-ALF: the method with \gaoc{the} \gaob{multilevel} perspective discrimination, \gaoc{the} context-aware learning block and \gaoc{the} global and local coherent discriminator (5) with TAL: the method without \gaoc{the} \gaob{multilevel} perspective discrimination and \gaoc{the} global and local coherent discriminator, \gaob{utilizing} \gaoc{the} traditional adversarial loss instead of the proposed adversarial loss. In addition, \gaoc{we compare the different loss components in Equation (\ref{L_loss}) to validate the effectiveness of the loss in HP-ALF}. (1) HP-ALF: the whole loss function in Equation (\ref{L_loss}); (2) without MPD: the loss function without $\mathcal{L}_{\mathrm{adv1}}$ (i.e., without \gaoc{the} \gaob{multilevel} perspective discrimination); (3) without LfMSE: the loss function without $\mathcal{L}_{\mathrm{fMSE}}$; (4) without Lvgg: the loss function without $\mathcal{L}_{\mathrm{VGG}}$; (5) with TAL: the loss function using \gaoc{the} traditional adversarial loss instead of the proposed adversarial loss in Equation (\ref{L_loss}); (6) without GLC: the loss function without $\mathcal{L}_{\mathrm{adv2}}$ (i.e., without the decoder of the discriminator).

\gaoc{Table \ref{Table_ha} shows the effectiveness of \gaob{the} current configuration in HP-ALF for all evaluation metrics. In particular, the values of PSNR in HP-ALF for the three datasets are 40.33, 37.45, 34.32.
The \gaob{FID values} for the Brain dataset, the Cardiac dataset and the Knee dataset are 55.75, 58.46, \gaob{and} 112.56, respectively.
These results indicate that the current configuration of HP-ALF achieves superior reconstruction performance.}

\gaoc{Figure \ref{Figure_loss}(a) shows that all loss functions increase quickly and converge after several epochs}. Moreover, HP-ALF with the whole loss function \gaob{enables faster} convergence with respect to other loss functions. For the loss function without the content loss, HP-ALF without $\mathcal{L}_{\mathrm{VGG}}$ and $\mathcal{L}_{\mathrm{fMSE}}$ \gaob{obtain} the third and fourth fastest convergence, respectively. For the loss function without the proposed adversarial loss, HP-ALF without GLC, HP-ALF without GLC and HP-ALF without MPD obtains the second, fifth and sixth fastest convergence, respectively. These results indicate that the proposed adversarial training loss (i.e., \gaob{the} combination of all above loss functions except traditional adversarial loss) achieves the fastest convergence.

\begin{figure}[t!]
\centering
	\includegraphics[width=0.49\textwidth]{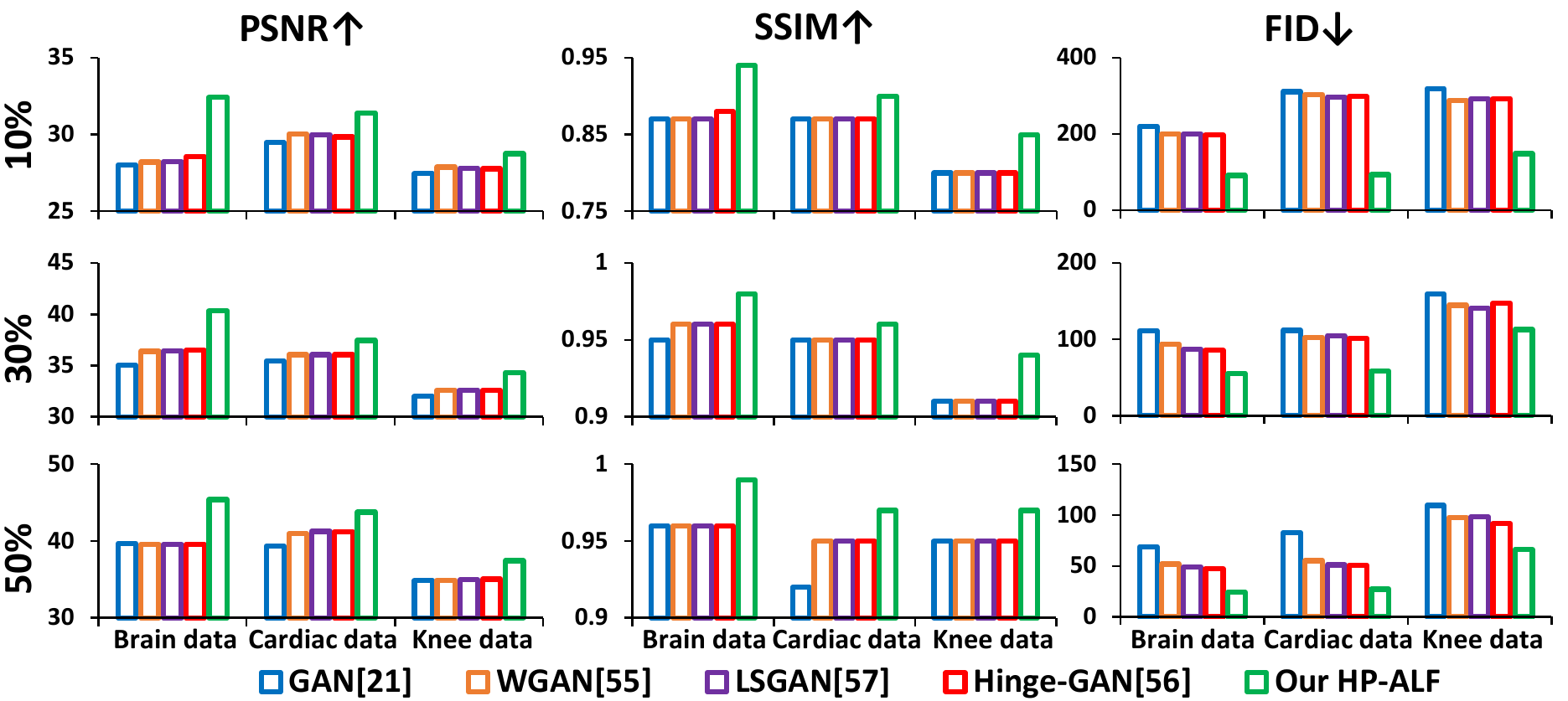}
\caption{The ablation study \gaob{for different} GAN objectives functions using 1D Gaussian mask. 10\%, 30\%, 50\% represent the percentages of the data sampling.}
\label{Table_GANs}
\end{figure}

\vspace{1.5mm}
\textbf{Performance of Multilevel Perspective Discrimination.}\label{Experiment for MPD}
\gaoc{This experiment includes four parts}. First, the training process is presented to validate the \gaob{convergence properties} for \gaob{multilevel} perspective discrimination in HP-ALF. Second, \gaoc{the} different objectives of GAN are \gaoc{compared} with \gaoc{the} \gaob{multilevel} perspective discrimination in HP-ALF, including the standard GAN (GAN) \cite{goodfellow2014generative}, WGAN \cite{1701.07875}, HingeGAN \cite{zhao2017energybased} \gaoc{and} LSGAN \cite{mao2017least}. Third, according to section \ref{sec:method RealnessGAN}, increasing the number of the \gaob{outcomes} $\mathbf{v}$ in $p_{\text {perspective }}$ can provide more information to the generator to improve the perceptual quality of the reconstruction results. To this end, \gaoc{the number of $\mathbf{v}$ is changed} while keeping the model unchanged to validate \gaob{that} \gaoc{the setting of this number} (i.e., the default number is 10) in HP-ALF is appropriate. \gaoc{Finally, the performance of the patch-level reduction of aliasing \gaob{artifacts} is visualized}. It shows the relationship by displaying the output of the discriminator (corresponding to $\mathbf{v}$) and the reconstruction error (corresponding to the image region). This is because each element in the output vector of the discriminator corresponds to an image attribute. The experiments set the different elements of this output as zero \gaob{to} make HP-ALF not focus on the reduction of aliasing \gaob{artifacts} in certain image regions.

Figure \ref{Figure_RealnessSamples} shows the convergence process \gaob{of HP-ALF}. \gaoc{The figure} \gaob{shows} that these curves correlate well with the perceptual quality of the generated samples. At the first 10000 iterations, the loss of the generator \gaoc{significantly decreases}, and the fine details in the generated samples (enlarged area in the figure) are sharpened. The loss of the generator converges in a small range in the subsequent process. These results indicate that \gaob{multilevel} perspective discrimination in HP-ALF can \gaob{reach optimality}.

Figure \ref{Table_GANs} presents the comparison results of \gaob{the} GAN baseline methods.
The results indicate that our HP-ALF with \gaob{multilevel} perspective discrimination obtains \gaoc{the} better scores in all the evaluation metrics compared to the GAN \gaoc{baseline methods}. Especially for the FID metric, the values for the \gaoc{Brain dataset}, the \gaoc{Cardiac dataset}, and the Knee dataset at \gaob{the} 10-fold undersampling rate are 90.79, 92.79 and 147.54, respectively.
\gaob{At 3.3$\times$ speed-up}, the \gaob{FID values} for the \gaoc{Brain dataset}, the \gaoc{Cardiac dataset}, and the Knee dataset are 55.75, 58.46, and 112.56, respectively.
In addition, the values for the \gaoc{Brain dataset}, the \gaoc{Cardiac dataset}, and the Knee dataset at 2$\times$ speed-up are 24.09, 27.04 and 66.16, respectively.
These results indicate that HP-ALF can achieve better reconstruction performance than other GAN baseline methods.

Figure \ref{Figure_loss}(b) shows that image quality improves as the number of $\mathbf{v}$ increases. The PSNR and SSIM values for HP-ALF with \gaob{multilevel} perspective discrimination are better than \gaob{those for} HP-ALF with TAL during the training process. The above results indicate that $\mathbf{v}$ can provide stronger guidance for $G$ to match the original image distribution. \gaoc{When the number of $\mathbf{v}$ exceeds 10, the image quality is not improved significantly}. Moreover, the convergence speed of the model is faster than others \gaob{when} the number of $\mathbf{v}$ is 10.

\begin{figure}
\centering
	\includegraphics[width=0.49\textwidth]{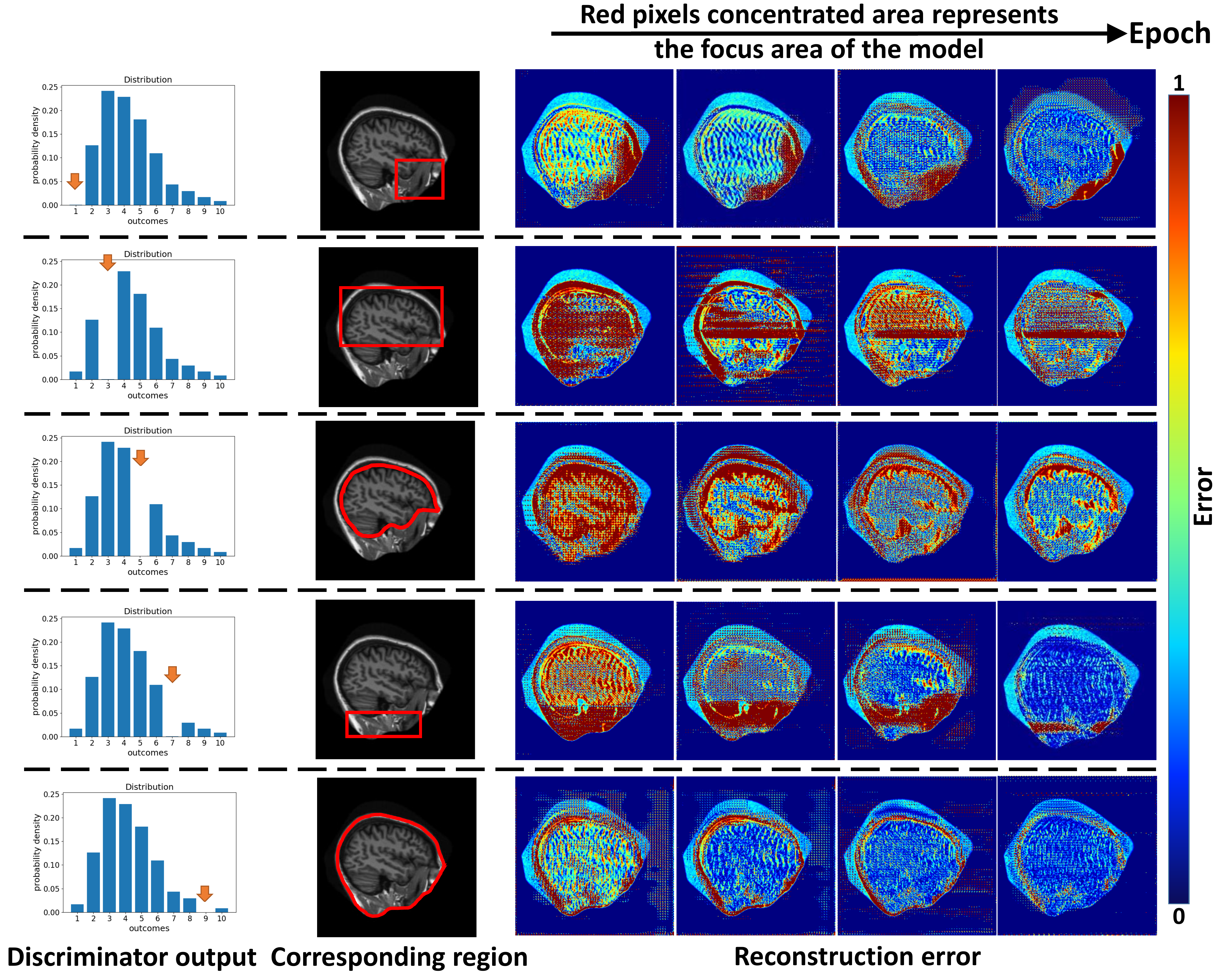}
\caption{The visual performance of the patch-level reduction of aliasing \gaob{artifacts}. It shows the relationship between the discriminator output and the reconstruction error. The experiment uses 1D Gaussian mask and is performed on 50\% data sampling. The \gaob{color} bar for the difference images is shown on the right. The changes of the selected elements (i.e., the discriminator output) lead to the changes of the image region (i.e., corresponding region). The aliasing \gaob{artifacts} in these image regions are not significantly reduced (i.e., reconstruction error).}
\label{Figure_distributionVal}
\end{figure}

Figure \ref{Figure_distributionVal} presents the visual performance of the patch-level reduction of aliasing \gaob{artifacts}. It shows that during the training procedure with the increase of epochs, the overall quality is gradually improved in the entire image. \gaoc{However, the region within the image also changes (where the aliasing \gaob{artifacts} are not significantly reduced), when the selected element (i.e., set to zero) of the output vector changes. Specifically, the image reconstruction at the edge of the brain \gaob{does not perform well} (i.e., has large error), when the selected element is located at the left or right side of the output vector (i.e., the first row and last row in the figure). Then, the image reconstruction at the \gaob{center} of the brain has large errors when the selected element is located at the peak of the output vector (i.e., the second row in the figure)}. Therefore, these results indicate \gaob{that different} $\mathbf{v}$ corresponds to the attributes in different image regions.

\vspace{1.5mm}
\textbf{Performance of Global and Local Coherent Discriminator.}
The experiments show the effectiveness of the discriminator by the influence of the per-pixel decision of the discriminator on the reconstruction error of the generator. The per-pixel decision investigates how the discriminator guides the generator to improve perceptual quality.

\begin{figure}
\centering
	\includegraphics[width=0.49\textwidth]{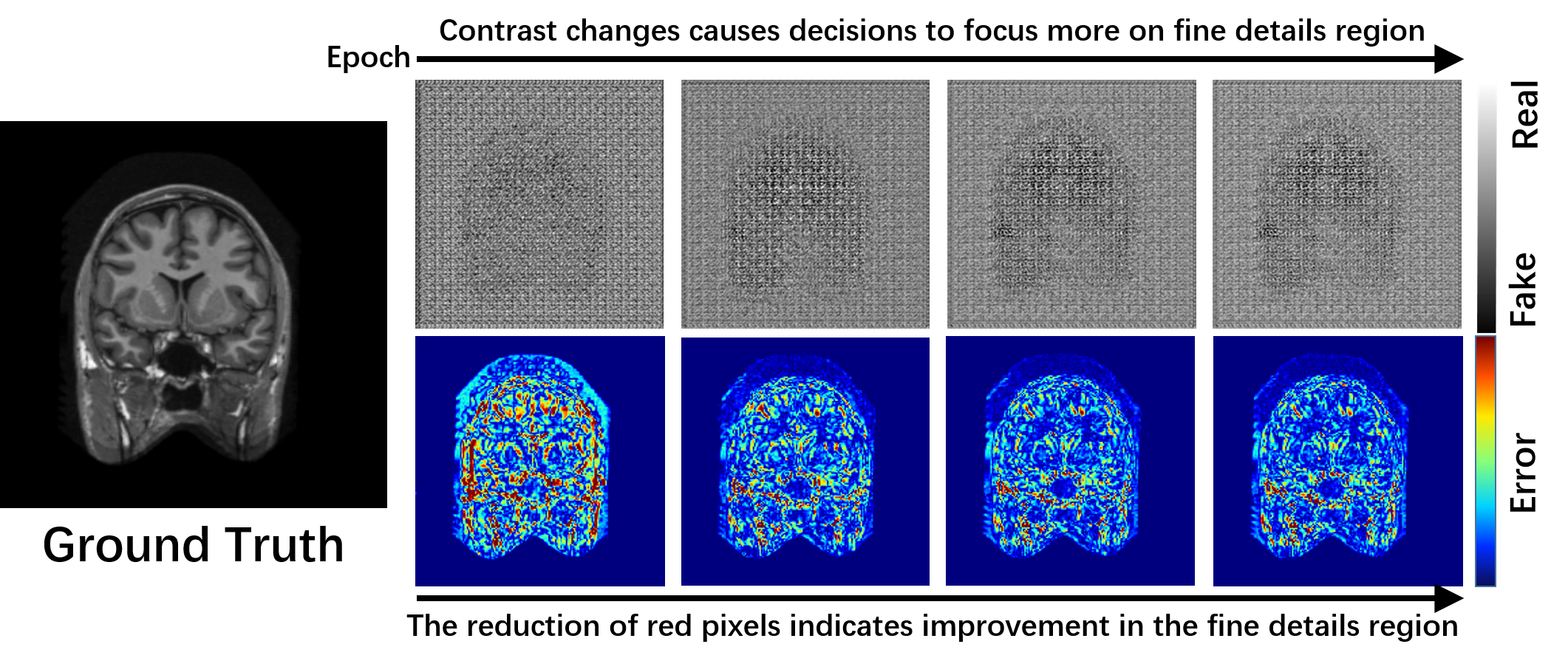}
\caption{Correlations between the per-pixel decision (top row) and the difference images (bottom row) investigate how the discriminator guides the generator. \gaob{Color} bars for the difference images and the feedback image are separately shown on the left and right. Darker \gaob{colors} in the per-pixel decision correspond to the discriminator confidence of pixel being fake, and then discriminator provides confidence to the generator to correct the fake regions (as shown in the difference images).}
\label{Figure_UnetGAN}
\end{figure}

Figure \ref{Figure_UnetGAN} shows that the background of the per-pixel decision (top row) first became bright, and then the region containing fine details became bright \gaoc{as the epoch increases}. Additionally, the output image becomes bright overall as the epoch increases. \gaob{The brighter (resp. darker) colors in the per-pixel decision correspond to the discriminator confidence of pixels being real (resp. being fake)}. \gaoc{These results indicate} that the background of the \gaob{reconstructed} image is recognized as fake by the discriminator. The focused area of the discriminator is gradually transferred from the image background to the region \gaoc{with} fine details. The per-pixel decision is correlated well with the difference (the bottom row \gaoc{in Figure \ref{Figure_UnetGAN}}) between \gaob{the reconstructed} image and ground truth. This indicates that the decision of the discriminator can guide the generator during the training process.

\begin{figure}
\centering
	\includegraphics[width=0.49\textwidth]{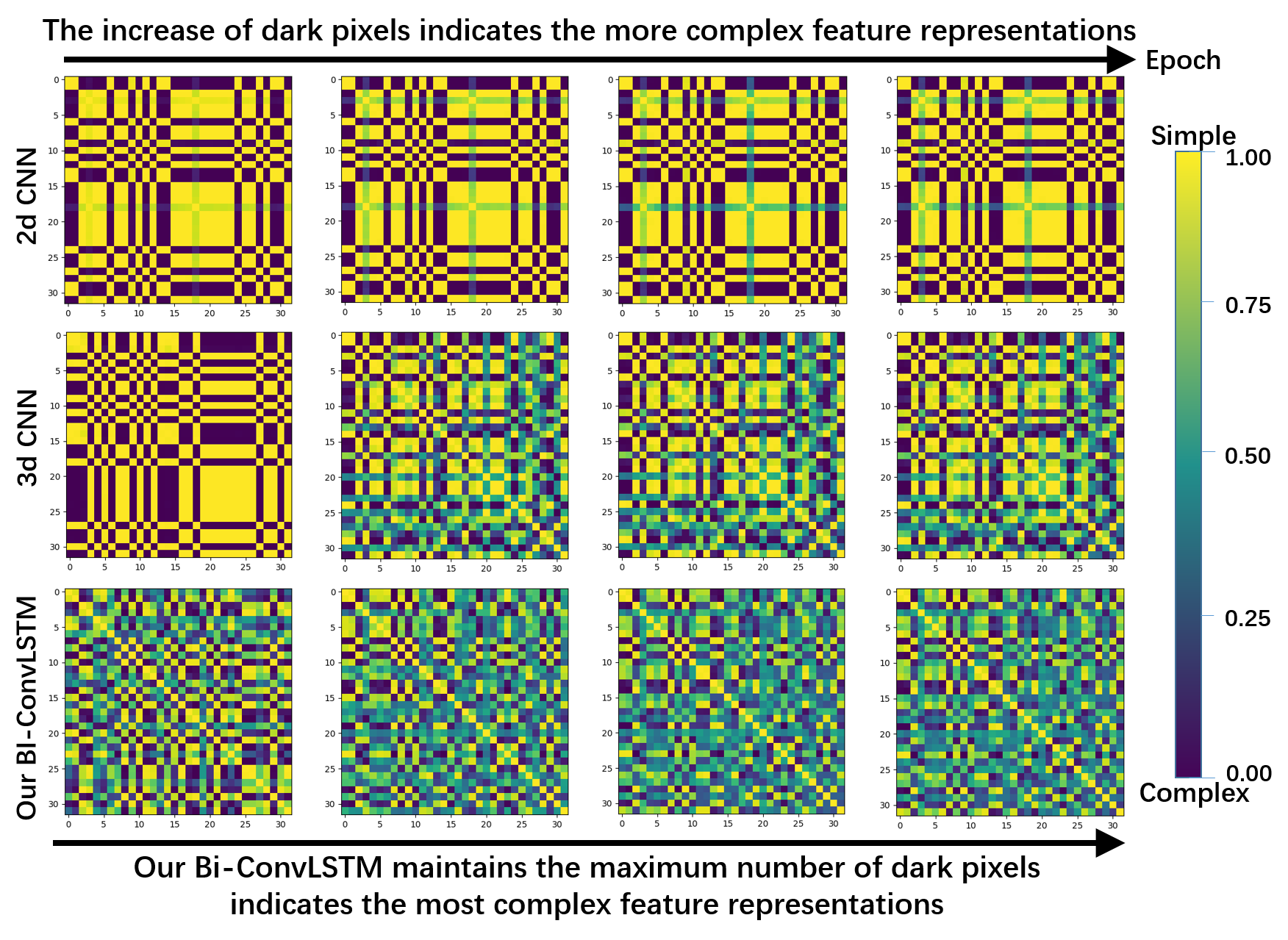}
\caption{The comparison of feature representations extracted by the different context-aware learning generators (U-net+2D CNN, U-net+3D CNN and U-net+Bi-ConvLSTM). The 2.5D U-net cannot the feature representations because it does not contain the block. The \gaob{color} of the pixel represents the similarity of the different features evaluated by cosine distance (darker \gaob{colors indicate} complex feature representations and brighter \gaob{indicate} simple representations). The increase of dark pixels shows that the feature representations of the module are gradually enriched. }
\label{Figure_LSTM}
\end{figure}

\vspace{1.5mm}
\textbf{Performance of Context-Aware Learning Generator.}
The experiments present the effectiveness of the proposed context-aware generator \gaoc{by considering} the kind \gaob{of generator}, the numbers of input slices, and the feature representation map.

Table~\ref{Table_LSTM} shows the comparison results of different context-aware learning generators. They \gaob{include the} proposed U-net+Bi-ConvLSTM (HP-ALF with Bi-ConvLSTM), U-net+2D CNN (HP-ALF with 2D CNN), U-net+3D CNN (HP-ALF with 3D CNN) and 2.5D U-net (HP-ALF with 2.5D U-net). Specifically, the 2.5D U-net \gaob{utilizes} the same U-net architecture \gaob{as those} in the comparative generators. The \gaoc{results} for HP-ALF with 3D CNN, HP-ALF with 2.5D U-net and HP-ALF with 2D CNN \gaoc{occupy} the second, third and fourth \gaob{places} in the most cases, respectively. However, HP-ALF with 2.5D U-net achieves the lowest scores \gaob{of} 28.36, 0.83, 166.12, \gaob{and} 0.82 in PSNR, SSIM, FID and PSIM, \gaob{respectively}, at 10$\times$ speed-up \gaoc{for the 5 input slices in the Knee dataset}.
These results indicate that HP-ALF with Bi-ConvLSTM achieves the \gaoc{best performance} \gaoc{with respect to other generators}.

Table \ref{Table_LSTM} also presents the comparison results using different numbers of input slices. The adjacent slices are considered as the input sequence. \gaoc{This table presents the effect of the number \gaob{of input} adjacent slices (set to 3$\sim$7) on different generators}. The results show that HP-ALF achieves the \gaoc{best performance} when the number of input adjacent slices is 5 for \gaoc{all comparative generators}. For example, when the dataset is brain at 10$\times$ acceleration, the PSNR, SSIM, FID and PSIM are 32.42, 0.94, 90.79, 0.90 for HP-ALF, 32.18, 0.94, 95.98, 0.90 for HP-ALF with 2.5D U-net, 31.74, 0.93, 94.90, \gaob{and} 0.90, \gaob{respectively}, for HP-ALF with 3D CNN and 31.67, 0.93, 97.81, \gaob{and} 0.88, \gaob{respectively}, for HP-ALF with 2D CNN, respectively. \gaoc{The best performance in the above results is reached when the input has five adjacent slices for all datasets}.

\begin{table*}[]
\centering
\caption{The ablation study with different context-aware learning generators (U-net+2D CNN, U-net+3D CNN, U-net+Bi-ConvLSTM and 2.5d U-net) using 1D Gaussian mask. HP-ALF with Bi-ConvLSTM is our method. 10\%, 30\%, 50\% represent the percentages of the data sampling. This study is performed under the different numbers of input adjacent slices (3, 4, 5, 6, 7).}
\resizebox{\textwidth}{!}{\renewcommand\arraystretch{0.85}{
\setlength{\tabcolsep}{2.5mm}{\begin{tabular}{l|l|l|c|l|l|l|l|l|l|l|l|l}
\hline
\multicolumn{13}{c}{\textbf{3 slice}}                                                                                                                                                                                                                                                                                                                                                                              \\
\hline
\multicolumn{1}{c|}{\multirow{3}{*}{\textbf{Comparison}}} & \multicolumn{12}{c}{\textbf{Brain data}}                                                                                                                                                                                                                                                                                                               \\
\cline{2-13}
\multicolumn{1}{c|}{}                                     & \multicolumn{4}{c|}{10\%}                                                                                                & \multicolumn{4}{c|}{30\%}                                                                                    & \multicolumn{4}{c}{50\%}                                                                                     \\
\cline{2-13}
\multicolumn{1}{c|}{}                                     & \multicolumn{1}{c|}{PSNR} & \multicolumn{1}{c|}{SSIM} & FID                                  & \multicolumn{1}{c|}{PSIM} & \multicolumn{1}{c|}{PSNR} & \multicolumn{1}{c|}{SSIM} & \multicolumn{1}{c|}{FID} & \multicolumn{1}{c|}{PSIM} & \multicolumn{1}{c|}{PSNR} & \multicolumn{1}{c|}{SSIM} & \multicolumn{1}{c|}{FID} & \multicolumn{1}{c}{PSIM}  \\
\hline
HP-ALF with 2D CNN                                        & 31.25                     & 0.93                      & 98.89                                & 0.86                      & 38.10                     & 0.98                      & 67.27                    & 0.97                      & 42.95                     & 0.99                      & 35.41                    & 0.99                      \\
HP-ALF with 3D CNN                                        & 31.59                     & 0.93                      & 96.49                                & 0.89                      & 39.82                     & 0.98                      & 63.49                    & 0.97                      & 43.89                     & 0.99                      & 32.71                    & 0.99                      \\
HP-ALF with 2.5D U-net                                    & 31.53                     & 0.93                      & 96.78                                & 0.89                      & 38.72                     & 0.98                      & 68.20                    & 0.97                      & 44.05                     & 0.99                      & 30.03                    & 0.99                      \\
HP-ALF with Bi-ConvLSTM                      & \textbf{32.20}            & \textbf{0.94}             & \textbf{91.56}                       & \textbf{0.90}             & \textbf{40.05}            & \textbf{0.98}             & \textbf{60.65}           & \textbf{0.98}             & \textbf{44.88}            & \textbf{0.99}             & \textbf{28.74}           & \textbf{0.99}             \\
\hline
\multicolumn{1}{c|}{\textbf{}}                            & \multicolumn{12}{c}{\textbf{Cardiac data}}                                                                                                                                                                                                                                                                                                             \\
\hline
HP-ALF with 2D CNN                                        & 30.33                     & 0.88                      & 120.79                               & 0.85                      & 36.11                     & 0.95                      & 81.19                    & 0.95                      & 39.20                     & 0.97                      & 70.98                    & 0.96                      \\
HP-ALF with 3D CNN                                        & 30.59                     & 0.89                      & 118.84                               & 0.87                      & 36.43                     & 0.95                      & 76.53                    & 0.96                      & 41.91                     & 0.97                      & 65.02                    & 0.96                      \\
HP-ALF with 2.5D U-net                                    & 30.78                     & 0.89                      & 99.00                                & 0.87                      & 36.09                     & 0.95                      & 79.50                    & 0.95                      & 40.05                     & 0.97                      & 69.84                    & 0.96                      \\
HP-ALF with Bi-ConvLSTM                      & \textbf{31.15}            & \textbf{0.90}             & \textbf{97.08}                       & \textbf{0.89}             & \textbf{36.74}            & \textbf{0.96}             & \textbf{72.36}           & \textbf{0.96}             & \textbf{42.67}            & \textbf{0.97}             & \textbf{53.68}           & \textbf{0.97}             \\
\hline
\multicolumn{1}{c|}{\textbf{}}                            & \multicolumn{12}{c}{\textbf{Knee data}}                                                                                                                                                                                                                                                                                                                \\
\hline
HP-ALF with 2D CNN                                        & 28.35                     & 0.83                      & 155.51                               & 0.82                      & 32.98                     & 0.91                      & 134.81                   & 0.94                      & 35.69                     & 0.95                      & 85.64                    & 0.92                      \\
HP-ALF with 3D CNN                                        & 28.46                     & 0.83                      & 151.64                               & 0.82                      & 33.17                     & 0.92                      & 131.83                   & 0.95                      & 36.21                     & 0.95                      & 79.85                    & 0.92                      \\
HP-ALF with 2.5D U-net                                    & 28.23                     & 0.82                      & 170.11                               & 0.82                      & 32.08                     & 0.90                      & 140.16                   & 0.95                      & 35.40                     & 0.95                      & 83.03                    & 0.92                      \\
HP-ALF with Bi-ConvLSTM                      & \textbf{28.55}            & \textbf{0.85}             & \textbf{149.68}                      & \textbf{0.82}             & \textbf{33.37}            & \textbf{0.92}             & \textbf{125.43}          & \textbf{0.96}             & \textbf{36.37}            & \textbf{0.96}             & \textbf{74.03}           & \textbf{0.93}             \\
\hline
\multicolumn{13}{c}{\textbf{4 slice}}                                                                                                                                                                                                                                                                                                                                                                              \\
\hline
\multicolumn{1}{c|}{\textbf{Comparison}}                  & \multicolumn{12}{c}{\textbf{Brain data}}                                                                                                                                                                                                                                                                                                               \\
\hline
HP-ALF with 2D CNN                                        & 31.48                     & 0.93                      & 98.74                                & 0.88                      & 38.72                     & 0.98                      & 65.02                    & 0.97                      & 42.12                     & 0.99                      & 35.69                    & 0.99                      \\
HP-ALF with 3D CNN                                        & 31.67                     & 0.93                      & 95.02                                & 0.90                      & 39.63                     & 0.98                      & 63.19                    & 0.97                      & 43.87                     & 0.99                      & 30.77                    & 0.99                      \\
HP-ALF with 2.5D U-net                                    & 31.89                     & 0.94                      & 96.55                                & 0.90                      & 39.40                     & 0.98                      & 66.67                    & 0.97                      & 43.89                     & 0.99                      & 31.02                    & 0.99                      \\
HP-ALF with Bi-ConvLSTM                      & \textbf{32.37}            & \textbf{0.94}             & \textbf{90.81}                       & \textbf{0.90}             & \textbf{40.20}            & \textbf{0.98}             & \textbf{58.41}           & \textbf{0.98}             & \textbf{45.11}            & \textbf{0.99}             & \textbf{25.99}           & \textbf{0.99}             \\
\hline
\multicolumn{1}{c|}{\textbf{}}                            & \multicolumn{12}{c}{\textbf{Cardiac data}}                                                                                                                                                                                                                                                                                                             \\
\hline
HP-ALF with 2D CNN                                        & 30.51                     & 0.88                      & 112.21                               & 0.85                      & 36.51                     & 0.95                      & 80.01                    & 0.95                      & 40.1                      & 0.97                      & 60.64                    & 0.96                      \\
HP-ALF with 3D CNN                                        & 30.51                     & 0.89                      & 106.5                                & 0.87                      & 36.88                     & 0.95                      & 69.90                    & 0.96                      & 42.49                     & 0.97                      & 49.76                    & 0.96                      \\
HP-ALF with 2.5D U-net                                    & 30.96                     & 0.89                      & 97.01                                & 0.88                      & 36.10                     & 0.95                      & 76.21                    & 0.95                      & 41.55                     & 0.97                      & 55.4                     & 0.96                      \\
HP-ALF with Bi-ConvLSTM                      & \textbf{31.31}            & \textbf{0.90}             & \textbf{95.68}                       & \textbf{0.89}             & \textbf{36.98}            & \textbf{0.96}             & \textbf{64.21}           & \textbf{0.96}             & \textbf{43.39}            & \textbf{0.98}             & \textbf{33.64}           & \textbf{0.97}             \\
\hline
\multicolumn{1}{c|}{\textbf{}}                            & \multicolumn{12}{c}{\textbf{Knee data}}                                                                                                                                                                                                                                                                                                                \\
\hline
HP-ALF with 2D CNN                                        & 28.37                     & 0.82                      & 150.19                               & 0.82                      & 33.10                     & 0.91                      & 130.71                   & 0.94                      & 36.50                     & 0.96                      & 77.96                    & 0.92                      \\
HP-ALF with 3D CNN                                        & 28.51                     & 0.82                      & 152.11                               & 0.82                      & 33.59                     & 0.92                      & 128.43                   & 0.95                      & 37.01                     & 0.96                      & 75.12                    & 0.92                      \\
HP-ALF with 2.5D U-net                                    & 28.26                     & 0.83                      & 162.00                               & 0.82                      & 32.88                     & 0.91                      & 131.11                   & 0.95                      & 36.12                     & 0.96                      & 81.11                    & 0.92                      \\
HP-ALF with Bi-ConvLSTM                      & \textbf{28.60}            & \textbf{0.85}             & \textbf{144.99}                      & \textbf{0.82}             & \textbf{34.20}            & \textbf{0.92}             & \textbf{111.36}          & \textbf{0.96}             & \textbf{37.11}            & \textbf{0.97}             & \textbf{73.10}           & \textbf{0.94}             \\
\hline
\multicolumn{13}{c}{\textbf{5 slice}}                                                                                                                                                                                                                                                                                                                                                                              \\
\hline
\multicolumn{1}{c|}{\textbf{Comparison}}                  & \multicolumn{12}{c}{\textbf{Brain data}}                                                                                                                                                                                                                                                                                                               \\
\hline
HP-ALF with 2D CNN                                        & 31.67                     & 0.93                      & 97.81                                & 0.88                      & 38.92                     & 0.98                      & 64.04                    & 0.97                      & 42.95                     & 0.99                      & 34.55                    & 0.99                      \\
HP-ALF with 3D CNN                                        & 31.74                     & 0.93                      & 94.90                                & 0.90                      & 39.52                     & 0.98                      & 61.43                    & 0.97                      & 43.89                     & 0.99                      & 28.19                    & 0.99                      \\
HP-ALF with 2.5D U-net                                    & 32.18                     & 0.94                      & 95.98                                & 0.90                      & 39.62                     & 0.98                      & 62.07                    & 0.98                      & 44.05                     & 0.99                      & 25.97                    & 0.99                      \\
HP-ALF with Bi-ConvLSTM                      & \textbf{32.42}            & \textbf{0.94}             & \textbf{90.79}                       & \textbf{0.90}             & \textbf{40.33}            & \textbf{0.98}             & \textbf{55.75}           & \textbf{0.98}             & \textbf{45.38}            & \textbf{0.99}             & \textbf{24.09}           & \textbf{0.99}             \\
\hline
\multicolumn{1}{c|}{\textbf{}}                            & \multicolumn{12}{c}{\textbf{Cardiac data}}                                                                                                                                                                                                                                                                                                             \\
\hline
HP-ALF with 2D CNN                                        & 30.57                     & 0.88                      & 120.79                               & 0.79                      & 36.61                     & 0.95                      & 76.71                    & 0.90                      & 40.76                     & 0.97                      & 34.83                    & 0.96                      \\
HP-ALF with 3D CNN                                        & 30.59                     & 0.89                      & 118.84                               & 0.79                      & 37.15                     & 0.95                      & 67.43                    & 0.91                      & 42.96                     & 0.98                      & 31.69                    & 0.96                      \\
HP-ALF with 2.5D U-net                                    & 31.12                     & 0.89                      & 96.30                                & 0.82                      & 36.37                     & 0.95                      & 73.08                    & 0.90                      & 42.46                     & 0.97                      & 33.28                    & 0.96                      \\
HP-ALF with Bi-ConvLSTM                      & \textbf{31.41}            & \textbf{0.90}             & \textbf{94.63}                       & \textbf{0.84}             & \textbf{37.45}            & \textbf{0.95}             & \textbf{58.46}           & \textbf{0.94}             & \textbf{43.77}            & \textbf{0.98}             & \textbf{27.04}           & \textbf{0.97}             \\
\hline
\multicolumn{1}{c|}{\textbf{}}                            & \multicolumn{12}{c}{\textbf{Knee data}}                                                                                                                                                                                                                                                                                                                \\
\hline
HP-ALF with 2D CNN                                        & 28.43                     & 0.82                      & 155.43                               & 0.82                      & 33.47                     & 0.92                      & 127.15                   & 0.97                      & 36.71                     & 0.97                      & 75.30                    & 0.92                      \\
HP-ALF with 3D CNN                                        & 28.59                     & 0.82                      & 153.20                               & 0.82                      & 34.11                     & 0.92                      & 123.82                   & 0.97                      & 37.14                     & 0.97                      & 72.82                    & 0.93                      \\
HP-ALF with 2.5D U-net                                    & 28.36                     & 0.83                      & 166.12                               & 0.82                      & 33.27                     & 0.92                      & 130.76                   & 0.97                      & 36.68                     & 0.97                      & 77.53                    & 0.92                      \\
HP-ALF with Bi-ConvLSTM                      & \textbf{28.75}            & \textbf{0.85}             & \textbf{147.54}                      & \textbf{0.82}             & \textbf{34.42}            & \textbf{0.94}             & \textbf{112.56}          & \textbf{0.97}             & \textbf{37.44}            & \textbf{0.97}             & \textbf{66.16}           & \textbf{0.94}             \\
\hline
\multicolumn{13}{c}{\textbf{6 slice}}                                                                                                                                                                                                                                                                                                                                                                              \\
\hline
\multicolumn{1}{c|}{\textbf{Comparison}}                  & \multicolumn{12}{c}{\textbf{Brain data}}                                                                                                                                                                                                                                                                                                               \\
\hline
HP-ALF with 2D CNN                                        & 31.33                     & 0.93                      & 97.49                                & 0.88                      & 38.77                     & 0.98                      & 66.74                    & 0.97                      & 42.9                      & 0.99                      & 37.77                    & 0.99                      \\
HP-ALF with 3D CNN                                        & 31.65                     & 0.93                      & 94.31                                & 0.90                      & 39.41                     & 0.98                      & 62.26                    & 0.97                      & 43.78                     & 0.99                      & 31.45                    & 0.99                      \\
HP-ALF with 2.5D U-net                                    & 32.02                     & 0.93                      & 92.36                                & 0.89                      & 39.52                     & 0.98                      & 63.11                    & 0.98                      & 44.08                     & 0.99                      & 28.11                    & 0.99                      \\
HP-ALF with Bi-ConvLSTM                      & \textbf{32.22}            & \textbf{0.94}             & \textbf{96.41}                       & \textbf{0.90}             & \textbf{40.25}            & \textbf{0.98}             & \textbf{58.88}           & \textbf{0.98}             & \textbf{45.24}            & \textbf{0.99}             & \textbf{28.45}           & \textbf{0.99}             \\
\hline
\multicolumn{1}{c|}{\textbf{}}                            & \multicolumn{12}{c}{\textbf{Cardiac data}}                                                                                                                                                                                                                                                                                                             \\
\hline
HP-ALF with 2D CNN                                        & 30.24                     & 0.89                      & 115.40                               & 0.85                      & 36.37                     & 0.95                      & 70.42                    & 0.95                      & 40.19                     & 0.97                      & 51.49                    & 0.96                      \\
HP-ALF with 3D CNN                                        & 30.51                     & 0.89                      & 108.89                               & 0.87                      & 37.05                     & 0.95                      & 68.76                    & 0.95                      & 42.60                     & 0.98                      & 53.12                    & 0.96                      \\
HP-ALF with 2.5D U-net                                    & 30.02                     & 0.89                      & 97.20                                & 0.87                      & 36.16                     & 0.95                      & 75.43                    & 0.95                      & 42.01                     & 0.97                      & 50.49                    & 0.96                      \\
HP-ALF with Bi-ConvLSTM                      & \textbf{31.30}            & \textbf{0.90}             & \textbf{96.44}                       & \textbf{0.89}             & \textbf{37.39}            & \textbf{0.96}             & \textbf{58.10}           & \textbf{0.96}             & \textbf{43.51}            & \textbf{0.98}             & \textbf{44.41}           & \textbf{0.97}             \\
\hline
\multicolumn{1}{c|}{\textbf{}}                            & \multicolumn{12}{c}{\textbf{Knee data}}                                                                                                                                                                                                                                                                                                                \\
\hline
HP-ALF with 2D CNN                                        & 28.31                     & 0.82                      & 159.74                               & 0.82                      & 33.29                     & 0.92                      & 119.96                   & 0.97                      & 36.60                     & 0.97                      & 74.33                    & 0.92                      \\
HP-ALF with 3D CNN                                        & 28.40                     & 0.82                      & 153.69                               & 0.82                      & 34.05                     & 0.92                      & 118.47                   & 0.97                      & 37.02                     & 0.97                      & 70.11                    & 0.93                      \\
HP-ALF with 2.5D U-net                                    & 28.29                     & 0.83                      & 155.48                               & 0.82                      & 33.25                     & 0.92                      & 121.21                   & 0.97                      & 36.43                     & 0.96                      & 77.90                    & 0.92                      \\
HP-ALF with Bi-ConvLSTM                        & \textbf{28.72}            & \textbf{0.85}             & \textbf{145.63}                      & \textbf{0.82}             & \textbf{34.29}            & \textbf{0.94}             & \textbf{115.49}          & \textbf{0.97}             & \textbf{37.30}            & \textbf{0.97}             & \textbf{68.63}           & \textbf{0.94}             \\
\hline
\multicolumn{13}{c}{\textbf{7 slice}}                                                                                                                                                                                                                                                                                                                                                                              \\
\hline
\multicolumn{1}{c|}{\textbf{Comparison}}                  & \multicolumn{12}{c}{\textbf{Brain data}}                                                                                                                                                                                                                                                                                                               \\
\hline
HP-ALF with 2D CNN                                        & 31.25                     & 0.93                      & \multicolumn{1}{l|}{101.22}          & 0.88                      & 38.74                     & 0.98                      & 72.91                    & 0.97                      & 42.63                     & 0.99                      & 39.51                    & 0.99                      \\
HP-ALF with 3D CNN                                        & 31.69                     & 0.93                      & \multicolumn{1}{l|}{97.55}           & 0.90                      & 39.36                     & 0.98                      & 65.79                    & 0.97                      & 43.77                     & 0.99                      & 32.67                    & 0.99                      \\
HP-ALF with 2.5D U-net                                    & 31.88                     & 0.93                      & \multicolumn{1}{l|}{96.71}           & 0.89                      & 39.50                     & 0.98                      & 67.81                    & 0.98                      & 44.01                     & 0.99                      & 28.92                    & 0.99                      \\
HP-ALF with Bi-ConvLSTM                      & \textbf{32.19}            & \textbf{0.94}             & \multicolumn{1}{l|}{\textbf{92.31}}  & \textbf{0.90}             & \textbf{40.21}            & \textbf{0.98}             & \textbf{61.73}           & \textbf{0.98}             & \textbf{45.19}            & \textbf{0.99}             & \textbf{27.06}           & \textbf{0.99}             \\
\hline
\multicolumn{1}{c|}{\textbf{}}                            & \multicolumn{12}{c}{\textbf{Cardiac data}}                                                                                                                                                                                                                                                                                                             \\
\hline
HP-ALF with 2D CNN                                        & 30.10                     & 0.89                      & \multicolumn{1}{l|}{118.70}          & 0.85                      & 36.28                     & 0.95                      & 84.36                    & 0.95                      & 40.07                     & 0.97                      & 59.96                    & 0.96                      \\
HP-ALF with 3D CNN                                        & 30.68                     & 0.89                      & \multicolumn{1}{l|}{110.44}          & 0.87                      & 37.02                     & 0.95                      & 71.28                    & 0.95                      & 42.40                     & 0.98                      & 55.69                    & 0.96                      \\
HP-ALF with 2.5D U-net                                    & 30.94                     & 0.89                      & \multicolumn{1}{l|}{99.40}           & 0.87                      & 36.05                     & 0.95                      & 77.79                    & 0.95                      & 41.22                     & 0.97                      & 57.25                    & 0.96                      \\
HP-ALF with Bi-ConvLSTM                      & \textbf{31.28}            & \textbf{0.90}             & \multicolumn{1}{l|}{\textbf{96.79}}  & \textbf{0.88}             & \textbf{37.31}            & \textbf{0.96}             & \textbf{58.98}           & \textbf{0.96}             & \textbf{43.24}            & \textbf{0.98}             & \textbf{51.44}           & \textbf{0.97}             \\
\hline
\multicolumn{1}{c|}{\textbf{}}                            & \multicolumn{12}{c}{\textbf{Knee data}}                                                                                                                                                                                                                                                                                                                \\
\hline
HP-ALF with 2D CNN                                        & 28.27                     & 0.82                      & \multicolumn{1}{l|}{163.79}          & 0.82                      & 33.23                     & 0.92                      & 127.67                   & 0.97                      & 36.54                     & 0.97                      & 77.80                    & 0.92                      \\
HP-ALF with 3D CNN                                        & 28.34                     & 0.82                      & \multicolumn{1}{l|}{159.88}          & 0.82                      & 34.01                     & 0.92                      & 125.43                   & 0.97                      & 36.98                     & 0.97                      & 75.01                    & 0.93                      \\
HP-ALF with 2.5D U-net                                    & 28.25                     & 0.83                      & \multicolumn{1}{l|}{169.48}          & 0.82                      & 33.21                     & 0.92                      & 129.77                   & 0.97                      & 36.25                     & 0.96                      & 80.66                    & 0.92                      \\
HP-ALF with Bi-ConvLSTM                       & \textbf{28.71}            & \textbf{0.85}             & \multicolumn{1}{l|}{\textbf{150.83}} & \textbf{0.82}             & \textbf{34.24}            & \textbf{0.94}             & \textbf{117.66}          & \textbf{0.97}             & \textbf{37.21}            & \textbf{0.97}             & \textbf{71.47}           & \textbf{0.94}             \\
\hline
\end{tabular}}}}
\label{Table_LSTM}
\end{table*}

\gaor{Figure \ref{Figure_LSTM} shows the feature representation ability of our Bi-ConvLSTM during the training process. It displays the similarity of the feature maps obtained by the context-aware learning block by the cosine distance $d(A, B)=A^{T} B /\|A\|\|B\|=\cos (\theta)$~\cite{qin2018convolutional}. If two feature maps are orthogonal, then $\cos (\theta)=0$ and if two feature maps are linearly correlated, then $\cos (\theta)=1$ (\gaoc{the} dark \gaob{colors} correspond to be orthogonal whereas bright \gaob{colors} correspond to \gaob{being} linearly correlated). When the values of the cosine distance are smaller, these feature maps are less similar. This means the feature maps learned from a context-aware learning block are diverse and complex. This experiment obtain 32 feature maps for each baseline context-aware learning block in the generator (i.e., 2D CNN, 3D CNN and Bi-ConvLSTM). The generator \gaoc{with} the 2.5D U-net \gaob{cannot} the similarity between the feature maps because it does not contain the context-aware learning block. The results show that our Bi-ConvLSTM has more dark pixels than the other baseline blocks. Especially \gaob{at epoch} 150, \gaoc{Bi-ConvLSTM obtains the most dark pixels. It also obtains more dark pixels than the other baseline blocks when the epoch is the same. Then, the mean values of the feature map for the Bi-ConvLSTM, 3D CNN and 2D CNN are 0.43, 0.53 and 0.66, respectively. These results indicate that the Bi-ConvLSTM from the context-aware learning generators learning rich feature information leads to the small average value and complex feature representation of the feature map.}}

\section{Conclusion}
In this paper, we propose \gaob{a} hierarchical perception adversarial learning framework to reconstruct \gaoc{high-quality MRI images}. This framework is implemented by the proposed GAN architecture, including the \gaob{multilevel} perspective discrimination, the global and local coherent discriminator and the context-aware learning generator. Specifically, the \gaob{multilevel} perspective discrimination \gaoc{provides} the information \gaob{for adversarial} learning from overall and regional perspectives. \gaoc{Then}, the global and local coherent discriminator \gaoc{enables} both global and local feedback to the generator. \gaoc{Finally}, the context-aware generator \gaob{builds relationships} among successive MRI slices to improve the reconstruction performance. \gaoc{Extensive experiments on three datasets (brain, heart and knee) show \gaob{the} effectiveness of our framework on high-quality MRI reconstruction}.

\begin{appendices}

\section{Details of multilevel perspective discrimination}\label{Appendix2}
The theoretical analysis for \gaob{multilevel} perspective discrimination is shown as follows:

\vspace{1.5mm}
\setcounter{theorem}{0}
\begin{theorem}
When $G$ is fixed, for any outcome $\mathbf{v}$ and input sample $x$, the optimal discriminator $D$ satisfies
\begin{linenomath*}
\begin{small}
\begin{equation}\nonumber
D_{G}^{\star}(\boldsymbol{x}, v)=\frac{ p_{\text {data }}(\boldsymbol{x})}{p_{\text {data }}(\boldsymbol{x})+p_{g}(\boldsymbol{x})}+
\frac{\mathcal{\mathcal { R }}_{1}(v) p_{\text {data }}(\boldsymbol{x})+\mathcal{R}_{0}(v) p_{g}(\boldsymbol{x})}{p_{\text {data }}(\boldsymbol{x})+p_{g}(\boldsymbol{x})}
\end{equation}
\end{small}
\end{linenomath*}
\end{theorem}

\vspace{1.5mm}
\emph{Proof}: Given a fixed $G$, the objective of $D$ is:
\begin{linenomath*}
\begin{small}
\begin{equation}\nonumber
\begin{aligned}
&\max _{G} \min _{D} V(G, D)= \mathbb{E}_{\boldsymbol{x} \sim p_{\text {data }}}[\mathcal{D}_{\mathrm{KL}}\left(\mathcal{R}_{1}(\mathbf{v}) \| D(\boldsymbol{x})\right)+\log (D(\boldsymbol{x}))] \\
&+\mathbb{E}_{\boldsymbol{x} \sim p_{g}}\left[\mathcal{D}_{\mathrm{KL}}\left(\mathcal{R}_{0}(\mathbf{v}) \| D(\boldsymbol{x})\right)+\log (1-D(\boldsymbol{x}))\right] \\
&=-\int_{\boldsymbol{x}}\left(p_{\text {data }}(\boldsymbol{x}) h\left(\mathcal{R}_{1}\right)+p_{g}(\boldsymbol{x}) h\left(\mathcal{R}_{0}\right)\right) d x \\
&-\int_{\boldsymbol{x}} \int_{v}\left(p_{\text {data }}(\boldsymbol{x}) \mathcal{R}_{1}(v)+p_{g}(\boldsymbol{x}) \mathcal{R}_{0}(v)\right) \log D(\boldsymbol{x}, v) d v d x,
\end{aligned}
\end{equation}
\end{small}
\end{linenomath*}
where $h\left(\mathcal{R}_{1}\right)$ and $h\left(\mathcal{R}_{0}\right)$ are their entropies. However, the first term as $C_{1}$ in the above equation is irrelevant to $D$, the objective thus is equivalent to
\begin{linenomath*}
\begin{equation}\nonumber
\begin{aligned}
&\min _{D} V(G, D)= -\int_{\boldsymbol{x}}\left(p_{\text {data }}(\boldsymbol{x})+p_{g}(\boldsymbol{x})\right) \\ &\int_{v} \frac{p_{\text {data }}(\boldsymbol{x}) \mathcal{R}_{1}(v)+p_{g}(\boldsymbol{x}) \mathcal{R}_{0}(v)}{p_{\text {data }}(\boldsymbol{x})+p_{g}(\boldsymbol{x})}\log D(\boldsymbol{x}, v) d v d x+C_{1} \\
&+\int_{\boldsymbol{x}}\left(p_{\text {data }}(\boldsymbol{x})+p_{g}(\boldsymbol{x})\right) \int_{v} \frac{p_{\text {data }}(\boldsymbol{x})}{p_{\text {data }}(\boldsymbol{x})+p_{g}(\boldsymbol{x})} \log D(\boldsymbol{x}, v) d v d x,
\end{aligned}
\end{equation}
\end{linenomath*}
\noindent where $p_{\boldsymbol{x}}(v)=\frac{p_{\text {data }}(\boldsymbol{x}) (\mathcal{R}_{1}(v)+1)+p_{g}(\boldsymbol{x}) \mathcal{R}_{0}(v)}{p_{\text {data }}(\boldsymbol{x})+p_{g}(\boldsymbol{x})}$ is a distribution defined on $\Omega$. The third term as $C_{2}$ is divided from the $C_{1}$. Let $C_{3}=p_{\text {data }}(\boldsymbol{x})+p_{g}(\boldsymbol{x})$,
we then have
\begin{linenomath*}
\begin{equation}\nonumber
\begin{aligned}
&\min _{D} V(G, D) = C_{1}+C_{2}\\
&+ \int_{\boldsymbol{x}} C_{3}(-\int_{v} p_{\boldsymbol{x}}(v) \log D(\boldsymbol{x}, v) d v+ h\left(p_{\boldsymbol{x}}\right)-h\left(p_{\boldsymbol{x}}\right)) d x \\
\end{aligned}
\end{equation}
\end{linenomath*}

For any valid $x$ in the above equation, when $\mathcal{D}_{\mathrm{KL}}\left(p_{\boldsymbol{x}} \| D(\boldsymbol{x})\right)$ achieves its minimum, $D$ obtains its optimal $D^{\star}$, leading to $D^{\star}(\boldsymbol{x})=\gaor{p_{\boldsymbol{x}}(v)}$, which concludes the proof.

\begin{theorem}
When $D=D_{G}^{\star}$, and there exists an outcome $v \in \Omega$ such that $\mathcal{R}_{1}(v) \neq \mathcal{R}_{0}(v)$, the maximum of $V\left(G, D_{G}^{\star}\right)$ is achieved if and only if $p_{g}=p_{\text {data }}$
\end{theorem}

\emph{Proof}: When $\gaor{p_{g}(\boldsymbol{x})}=\gaor{p_{\text{data}}(\boldsymbol{x})}$, $D_{G}^{\star}(\boldsymbol{x}, v)=\frac{(\mathcal{R}_{1}(v)+1)+\mathcal{R}_{0}(v)}{2}$, we have
\begin{linenomath*}
\begin{equation}\nonumber
\begin{aligned}
V^{\star}\left(G, D_{G}^{\star}\right)= \mathcal{D}_{\mathrm{KL}}\left(\mathcal{R}_{1} \| D^{\star}(\boldsymbol{x})\right) + \mathcal{D}_{\mathrm{KL}}\left(\mathcal{R}_{0} \| D^{\star}(\boldsymbol{x})\right).
\end{aligned}
\end{equation}
\end{linenomath*}

Subtracting $V^{\star}\left(G, D_{G}^{\star}\right)$ from $V\left(G, D_{G}^{\star}\right)$ gives

\begin{linenomath*}
\begin{equation}\nonumber
\begin{aligned}
V^{\prime}\left(G, D_{G}^{\star}\right)=&V\left(G, D_{G}^{\star}\right)-V^{\star}\left(G, D_{G}^{\star}\right)
\\
=&-2 \mathcal{D}_{\mathrm{KL}}\Big(
\frac{\gaor{p_{\mathrm{data}}(\boldsymbol{x})}\mathcal{R}_{1}+ \gaor{p_{g}(\boldsymbol{x})}\mathcal{R}_{0}}{2}
\\
&\| \frac{\left(\gaor{p_{\mathrm{data}}(\boldsymbol{x})}+\gaor{p_{g}(\boldsymbol{x})}\right)\left(\mathcal{R}_{1}+\mathcal{R}_{0}-1\right)\frac{1}{2}}{4}
\Big)
\end{aligned}
\end{equation}
\end{linenomath*}


Since $V^{\star}\left(G, D_{G}^{\star}\right)$ is a constant with respect to G, maximising $V\left(G, D_{G}^{\star}\right)$ is equivalent to maximising $V^{\prime}\left(G, D_{G}^{\star}\right)$. The optimal $V^{\prime}\left(G, D_{G}^{\star}\right)$ is achieved if and only if the KL divergence reaches its minimum, where
\begin{linenomath*}
\begin{equation}\nonumber
\begin{aligned}
\frac{\gaor{p_{\mathrm{data}}(\boldsymbol{x})}
 \mathcal{R}_{1}+\gaor{p_{g}(\boldsymbol{x})} \mathcal{R}_{0}}{2} =\frac{\left(\gaor{p_{\mathrm{data}}(\boldsymbol{x})}+\gaor{p_{g}(\boldsymbol{x})}\right)\left(\mathcal{R}_{1}+\mathcal{R}_{0}\right)\frac{1}{2}}{4}
\end{aligned}
\end{equation}
\end{linenomath*}
\begin{linenomath*}
\begin{equation}\nonumber
\begin{aligned}
\left(\gaor{p_{\mathrm{data}}(\boldsymbol{x})}-\gaor{p_{g}(\boldsymbol{x})}\right)\left(\mathcal{R}_{1}-\mathcal{R}_{0}\right)\frac{1}{2} =0
\end{aligned}
\end{equation}
\end{linenomath*}
for any valid $\boldsymbol{x}$ and $\mathbf{v}$. Hence, as long as there exists a valid $\mathbf{v}$ that $\mathcal{R}_{1}(v) \neq \mathcal{R}_{0}(v)$, we have \gaor{$p_{\mathrm{data}}(\boldsymbol{x})=p_{g}(\boldsymbol{x})$} for any valid $\boldsymbol{x}$.
If one views the above equation as a cost function to minimise, when \gaor{$p_{\mathrm{data}}(\boldsymbol{x}) \neq p_{g}(\boldsymbol{x})$}, the larger the difference between $\mathcal{R}_{1}(v)$ and $\mathcal{R}_{0}(v)$ is, the stronger the constraint on $G$ becomes.

\end{appendices}

\bibliographystyle{IEEEtran}
\bibliography{ref}

\end{document}